\documentclass[twocolumn,portrait]{aastex62}

\usepackage{natbib}
\usepackage[utf8]{inputenc}
\usepackage{newunicodechar}
\usepackage{booktabs}
\usepackage{multirow}
\usepackage{longtable}
\usepackage{array}
\usepackage{amsmath}
\usepackage{rotating}
\usepackage{soul}

\shortauthors{Gootkin et al.}
\shorttitle{}

\begin{document}

\title{A New Catalog of 100,000 Variable \emph{TESS} A-F Stars Reveals a Correlation Between $\delta$ Scuti Pulsator Fraction and Stellar Rotation}

\author[0000-0003-0922-138X]{Keyan Gootkin}
\affiliation{Institute for Astronomy, University of Hawai‘i, 2680 Woodlawn Drive, Honolulu, HI 96822, USA}
\author[0000-0003-2400-6960]{Marc Hon}
\affiliation{Institute for Astronomy, University of Hawai‘i, 2680 Woodlawn Drive, Honolulu, HI 96822, USA}
\affiliation{Kavli Institute for Astrophysics and Space Research, Massachusetts Institute of Technology,
77 Massachusetts Avenue, Cambridge, MA 02139, USA}
\author[0000-0001-8832-4488]{Daniel Huber}
\affiliation{Institute for Astronomy, University of Hawai‘i, 2680 Woodlawn Drive, Honolulu, HI 96822, USA}
\affiliation{Sydney Institute for Astronomy, School of Physics, University of Sydney NSW 2006, Australia}
\author[0000-0003-3244-5357]{Daniel R.\ Hey}
\affiliation{Institute for Astronomy, University of Hawai‘i, 2680 Woodlawn Drive, Honolulu, HI 96822, USA}
\author[0000-0001-5222-4661]{Timothy R. Bedding}
\affiliation{Sydney Institute for Astronomy, School of Physics, University of Sydney NSW 2006, Australia}
\author[0000-0002-5648-3107]{Simon J. Murphy}
\affiliation{Centre for Astrophysics, University of Southern Queensland, Toowoomba, QLD 4350, Australia}

\correspondingauthor{Keyan Gootkin}
\email{gootkin@hawaii.edu}
\begin{abstract}
$\delta$ Scuti variables are found at the intersection of the classical instability strip and the main sequence on the Hertzsprung-Russell diagram.
With space-based photometry providing millions of light-curves of A-F type stars, we can now probe the occurrence rate of $\delta$ Scuti pulsations in detail.
Using 30-min cadence light-curves from NASA’s Transiting Exoplanet Survey Satellite’s (TESS) first 26 sectors, we identify variability in 103,810 stars within 5-24 cycles per day down to a magnitude of $T=11.25$.
We fit the period-luminosity relation of the fundamental radial mode for $\delta$ Scuti stars in the \textit{Gaia} $G$-band, allowing us to distinguish classical pulsators from contaminants for a subset of 39,367 stars.
Out of this subset, over \textbf{15,918} are found on or above the expected period-luminosity relation.
We derive an empirical red edge to the classical instability strip using \textit{Gaia} photometry.
The center where pulsator fraction peaks at 50-70\%, combined with the red edge, agree well with previous work in the \textit{Kepler} field.
While many variable sources are found below the period-luminosity relation, over 85\% of sources inside of the classical instability strip derived in this work are consistent with being $\delta$ Scuti stars.
The remaining 15\% of variables within the instability strip are likely hybrid or $\gamma$ Doradus pulsators.
Finally, we discover strong evidence for a correlation between pulsator fraction and spectral line broadening from the Radial Velocity Spectrometer (RVS) aboard the \textit{Gaia} spacecraft, confirming that rotation has a role in driving pulsations in $\delta$ Scuti stars.

\end{abstract}

\section{Introduction}
$\delta$ Scuti variables are stars of spectral type A0-F5 on or near the main sequence and within the classical instability strip, with luminosities of roughly $2-50 L_\sun$ and masses of roughly $1.5-2.3 M_\sun$ \citep{BregerReview1979,GoupilReview2005,HandlerReview2009,GuzikReview2021,KurtzReview2022}.

In theory, any star inside of the instability strip should have the partial ionization layers which drive $\delta$ Scuti pulsations through the $\kappa$-mechanism \citep{DupretTheoreticalInstabilityStrip,DupretInstabilityStrip2}. \citet[][]{KeplerDeltaScutiPulsatorFraction} used \textit{Kepler} data to detect pulsations in 1988 stars and showed that the fraction of stars which are pulsators peaks at only 70\% in the center of the instability strip using a sample of over 15,000 A-F stars in the \textit{Kepler} field. While the distribution of observed pulsation frequencies in many $\delta$ Scuti stars have been reported to correlate with the stars' fundamental properties \citep[e.g. $\nu_{\mathrm{max}}$ with $T_{\mathrm{eff}}$; ][]{Balona2011,Barcelo2018,BowmanKeplerdsct2018,Hasanzadeh2021}, theoretical progress has not explained the basic question of which stars should or should not pulsate \citep{KeplerDeltaScutiPulsatorFraction,BalonaModeSelection,TESSPleadies}.

Because of the $\kappa$-mechanism's reliance on the presence of helium at a particular depth within a star, this mechanism must be affected by the chemical structure of a star \citep{GuzikOpacity2018}. Chemically peculiar stars such as the metallically lined A-stars (Am stars) have very low pulsator fractions \citep{AmPulse1970,AmPulse1977,GuzicAmStars2021}. Am stars are notably slow rotators, which is thought to suppress the $\kappa$-mechanism via gravitational settling of helium out of the ionization zone which drives pulsation \citep{dsctRotation1974,dsctRotation1980,dsctRot2015}. This diffusion processes has been invoked to explain observations that $\delta$ Scuti stars tend to be moderate or rapid rotators \citep{dsctRotation1997,dsctRotation2009}. However, these studies were limited by small sample sizes of 10s to 100s of stars.

A challenge with using \textit{Kepler} data to investigate the occurrence rate of $\delta$ Scuti stars is its complex selection function, which focused on solar-type stars in order to detected transiting exoplanets \citep{KeplerSelection2010,KeplerSelection2021}. Such a selection function---the heuristics by which a survey selects targets to observe---means that the \textit{Kepler} sample of A-F stars is not necessarily representative of the larger galactic populations. A larger sample of $\delta$ Scuti variables that is only limited in magnitude will be valuable to verify the instability strip and to examine trends in pulsation properties (i.e. frequency and amplitude) with other stellar quantities. This paper will expand upon the work done in the \textit{Kepler} field by using NASA’s Transiting Exoplanet Survey Satellite \citep[\emph{TESS};][]{TESS}. Because \emph{TESS} surveys the entire sky without a particular selection function, it allows the most expansive investigation of $\delta$ Scuti pulsators conducted to date \citep{Antoci2019,Balona2020,BaracPLR,Skarka2022,Xue2023,Read2024}. 


\begin{figure}[ht!] \label{fig:cmd}
    \plotone{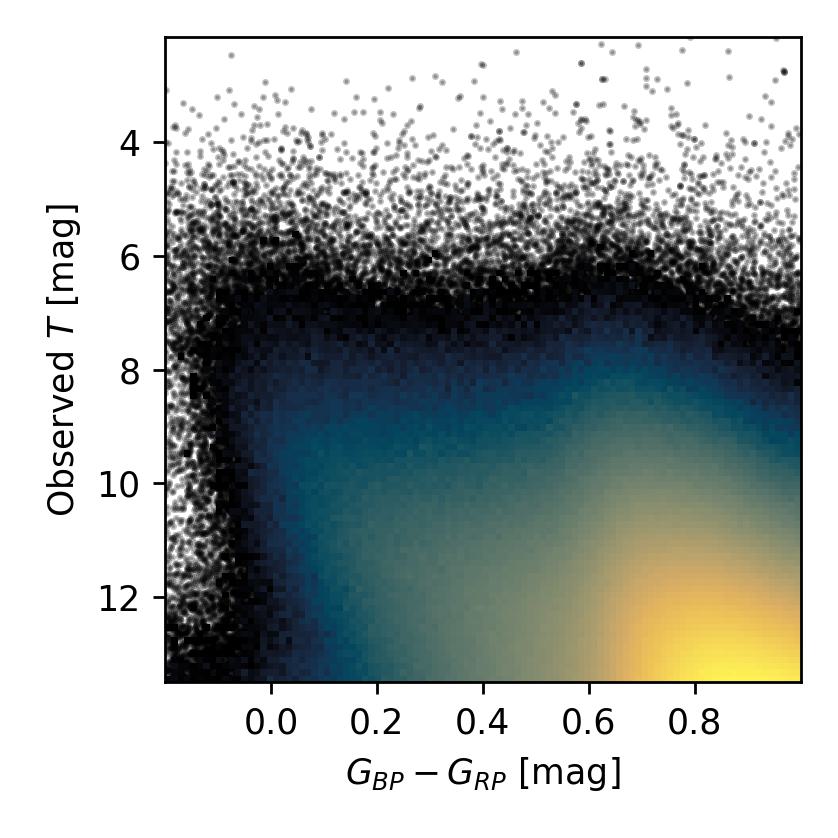}
    \caption{A 2D histogram showing the density of all 6,884,170 targets in our sample shown in color versus apparent \emph{TESS}-magnitude. For bins (where each bin is 0.012 mag $\times$ 0.115 mag in size) without at least 10 objects we plot the positions of stars as a scatter plot. These quantities have not been corrected for reddening or extinction.}
\end{figure}

\section{Observations \& Target Selection} \label{sec:obs}
\subsection{Sample Selection} \label{sec:sampleselction}
We initially construct our sample from the \emph{TESS} Input Catalog \citep[TIC; ][]{TIC} based on the \textit{Gaia} color, $G_{BP}-G_{RP}$, tabulated from Gaia Data Release 3 \citep{Gaia1,Gaia2}. We choose bounds which surround the classical instability strip near to the main-sequence ($-0.2 < G_{BP}-G_{RP} < 1$) according to synthetic photometry from MIST models of main-sequence, solar metallicity stars \citep[MESA (Modules for Experiments in Stellar Astrophysics) Isochrones \& Stellar Tracks;][]{MESA2011,MESA2013,MESA2015,MISTa,MISTb}. There are 6,884,170 objects which fit the constraints $-0.2 < G_{BP}-G_{RP} < 1$ and $T<13.5$. The distribution of this sample in color and apparent \emph{TESS} magnitude is shown in Figure \ref{fig:cmd}.  Because we choose a volume limited dust map (\S\ref{sec:dust}) we limit our analysis to only the brightest 1 million objects in this work.

\subsection{\emph{TESS} Light-Curves} \label{sec:lightcurves}
We use Quick-Look Pipeline light-curves \citep[QLP;][]{QLP1,QLP2,QLPextended}, which were generated from \emph{TESS}'s 30-minute cadence Full-Frame Images (FFIs), for sectors 1-26. The QLP has published light-curves for every observed target in the \emph{TESS} FFIs with $T<13.5$. From each light-curve we extracted the time, KSPSAP flux which has slow trends and systematics removed, flux error, and quality columns. We selected only timestamps with a quality flag of 0, the strictest quality standard available. From the 1 million brightest objects we are able to analyze such light-curves for 754,909 sources, we refer to these sources as our \textit{Processed Sample} (defined in \S\ref{sec:samples}). The remaining sources either were not observed in sectors 1-26, or did not have data with a quality flag of 0.

Each sector was analyzed separately, and not combined with other sectors for the same star. We do this to ensure a uniform treatment of data, irrespective of location on the sky and therefore number of sectors observed. Where multiple sectors of data are available, we use data from the sector with the most statistically significant result (as described in \S\ref{sec:methods}, and include the sector utilized in Table \ref{table}).

\subsection{Interstellar Extinction} \label{sec:dust}

We correct for extinction using the 3D dust map from \citet{LiekeDustMap}, via the Python package \texttt{dustmaps} \citep{dustmaps}. This dust map was chosen based on the coverage and reliability. \textbf{The \citet{LiekeDustMap} dust map is defined in a 740 pc × 740 pc × 540 pc box centered on the Sun, ensuring that the closest, apparently brightest stars can be analyzed}. Compared to Gaia derived values of $A_G$, this dust map covers a larger portion of the sample described above. Additionally, for stars with $A_G < 0.25$ according to the \citet{LiekeDustMap} dust map, several thousand stars had Gaia $A_G$ values from 1 to 6 magnitudes. Such large extinction values for nearby stars are unrealistic, and we thus assume the \citet{LiekeDustMap} values to be more reliable for these stars. 

For each of the targets we integrate the extinction density along the line of sight towards the target\textbf{, using distance values provided by Gaia DR3}. For the reddening---$E(G_{BP}-G_{RP})$---we estimate the extinction in $G_{BP}$ and $G_{RP}$ by fitting the extinction law from \citet{F99} to the extinction in $G$. Using this method we can correct for extinction for 56.9\% of the stars which had at least one sector of \emph{TESS} data, or nearly 430,000 targets (the \textit{Dust Corrected} sample as described in \S\ref{sec:samples}). Since the \citet{LiekeDustMap} dust map is defined in a 740 pc × 740 pc × 540 pc box centered on the Sun, this method biases our \textit{Dust Corrected} sample towards the nearest A-F type stars, but maintains full-sky coverage.

\begin{figure*}[ht!]
    \centering
    \plotone{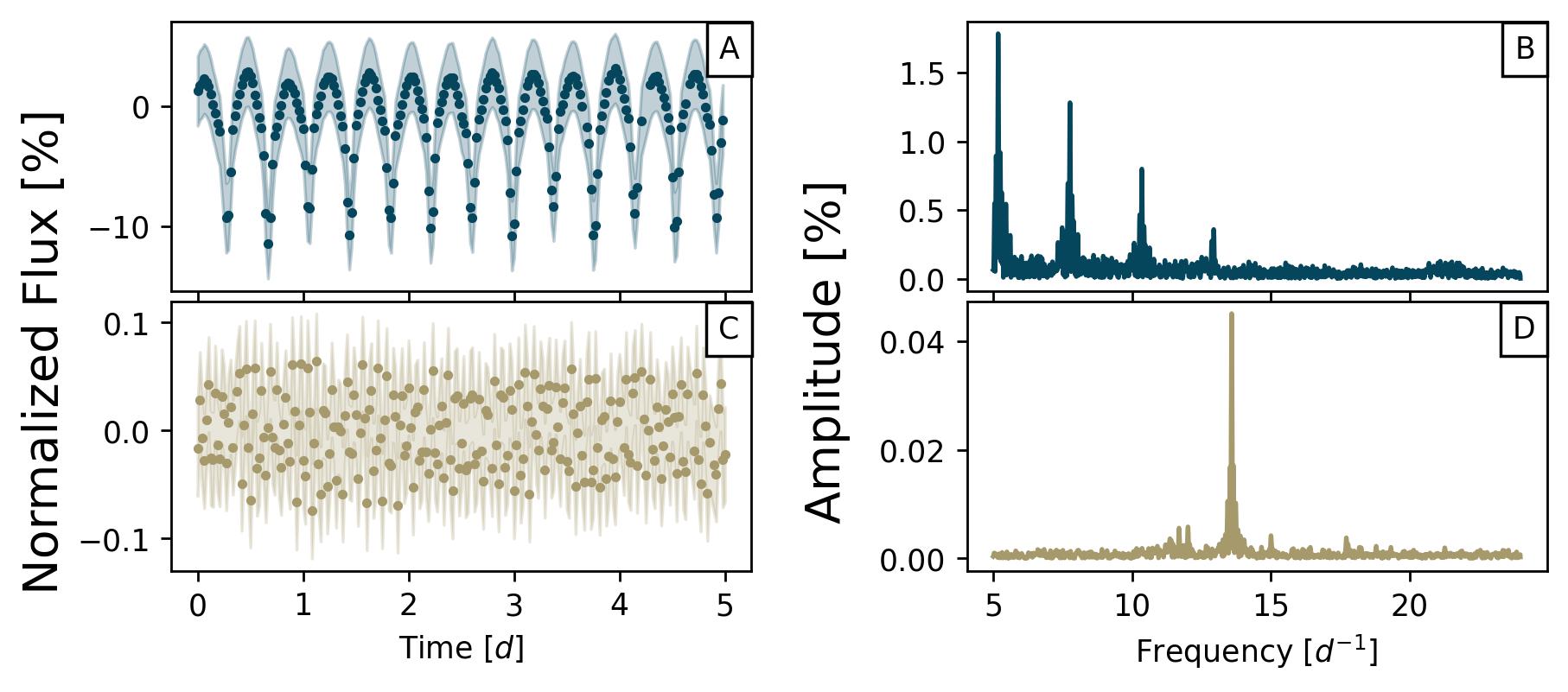}
    \caption{Light-curves (panels A and C; shaded regions represent errors) and amplitude spectra (panels B and D) of the known eclipsing binary TIC 14842303 in sector 14 \citep[TYC 2697-130-1; panels A and B;][]{VillanovaEB} and the $\delta$ Scuti variable star TIC 67991192 ($\gamma$ Boo; panels C and D) in sector 16. These are representative examples of their respective categories, and illustrate the differences which allow for the identification of pulsators as described in \S\ref{sec:class_eb}.}
    \label{fig:dsctvseblc}
\end{figure*}

\begin{figure}
    \centering
    \plotone{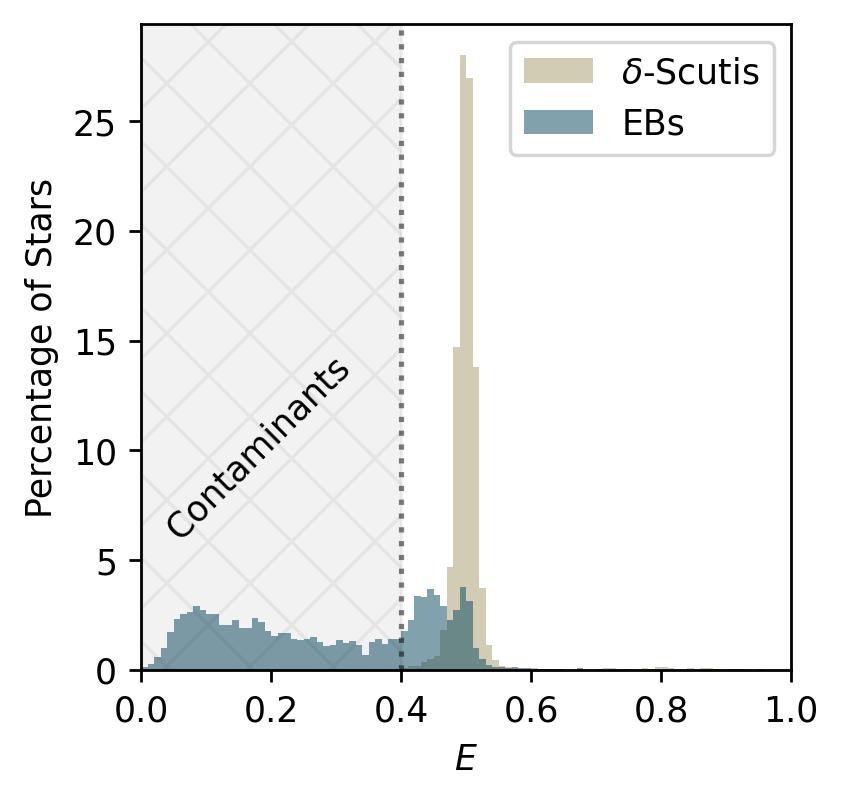}
    \caption{Histograms of $E$-parameters for \citet{KeplerDeltaScutiPulsatorFraction}'s sample of $\delta$ Scuti variables in gold and a sample of eclipsing binaries from \citet{VillanovaEB} in dark blue. The vertical axis is constructed such that each bin height represents the percentage of stars from that sample in that bin. The grey hatched region shows the range of values which would result in marking an object an \textit{Eclipsing Binary} in table \ref{table}. Both distributions are partially transparent, such that where the two distributions overlap about $E=0.5$ and Count Density$<4\%$, there is a darker color.}
    \label{fig:dsctebhist}
\end{figure}

\section{Methodology} \label{sec:methods}
\subsection{Amplitude Spectrum Calculation} \label{sec:spec}
For each of the light-curves---created from each sector of observations of each star---we calculate a Lomb-Scargle periodogram \citep{Lomb,Scargle,vanderplas} using \texttt{AstroPy}'s \texttt{LombScargle} method \citep{astropy2013,astropy2018,Astropyv5}.The periodograms presented in this work were calculated between 5 and 24 cycles per day ($d^{-1}$). The lower limit of 5 $d^{-1}$ is chosen to exclude low frequency pulsations that are characteristic of other types of pulsators, such as $\gamma$ Doradus variables \citep{dsct_vs_gdor2002,dsct_hybrids_corot}. It is worth noting, however, that this might remove the highest-luminosity $\delta$ Scuti stars, which can have frequencies below this cutoff \citep[see Fig. 2 in ][]{BaracPLR}. A lower threshold of 1 $d^{-1}$ was explored, however, below 5 $d^{-1}$ our selection (Figure \ref{fig:plr}) becomes dominated by a large number of non-$\delta$ Scuti pulsators in the range 1-5 $d^{-1}$. The upper limit---24 $d^{-1}$---is the Nyquist frequency for 30-minute cadence \emph{TESS} FFI observations. This upper limit will significantly limit the ability to detect the highest frequency $\delta$ Scuti pulsators \citep[e.g. Figure 9 of ][]{Hey2021}, which will only be detected by aliases of the true pulsation frequency mirrored across the Nyquist frequency \citep{FastPulsatingDeltaScutiNature}. The \emph{TESS} sample of $\delta$ Scuti stars in \citet{Read2024} shows that a significant fraction of $\delta$ Scuti stars pulsate more rapidly than 24 $d^{-1}$. In their sample of 851 $\delta$ Scuti stars $\sim 46\%$ have recorded frequencies greater than 24 $d^{-1}$. Assuming that this fraction remains constant across the instability strip, we expect a similar fraction of $\delta$ Scuti stars in this work to be affected by this aliasing.

\subsection{Variability Identification} \label{sec:classification}
We use the method described in \citet{falsealarmprobs} to estimate the false alarm probability (FAP) for peaks in our amplitude spectra. FAP is the probability that white noise could produce a single peak of a given amplitude. For the purposes of this paper we define our sample of \textit{Variable Sources} as the objects for which we detect at least one peak in the periodogram between 5 and 24 $d^{-1}$ that has a FAP less than 1\%. Most other variable sources within our color cuts (slowly pulsating B-stars, RR-Lyrae variables, Cepheid variables) pulsate at frequencies below 5 $d^{-1}$ \citep{Gaiadr3hotvariability}, so variables which vary more rapidly than 5 $d^{-1}$ are likely $\delta$ Scuti variables. Reddened $\beta$-Cephei variables, which pulsate on scales of hours \citep{BetaCephieds}, may still remain, however these objects are rare compared to $\delta$-Scuti stars and should be excluded from the \textit{Dust-Corrected} sample as described in \S\ref{sec:classification}.

\subsection{Identification of False Positives} \label{sec:class_eb}
\subsubsection{Eclipsing Binaries}
We expect the largest fraction of false positives will come from eclipsing binaries (EBs). While most EBs are periodic at frequencies far lower than 5 $d^{-1}$, the highly non-sinusoidal shape of eclipses causes peaks in the periodogram at many integer multiples of the orbital frequency (referred to as harmonic peaks). These harmonic peaks may stretch into the frequency range we consider, meaning that harmonics can cause false-positives \citep[][]{Balona2011,KeplerDeltaScutiPulsatorFraction,Read2024}. The same argument can be made for ellipsoidal variables/contact binaries, however EBs exhibit significantly larger deviations from sinusoidal signals. We will thus focus on detecting EBs in particular, and instead exclude ellipsoidal/contact binaries through the $\delta$-Scuti period-luminosity relation (\S\ref{sec:plr}).

\begin{figure}[ht!]
    \centering
    \plotone{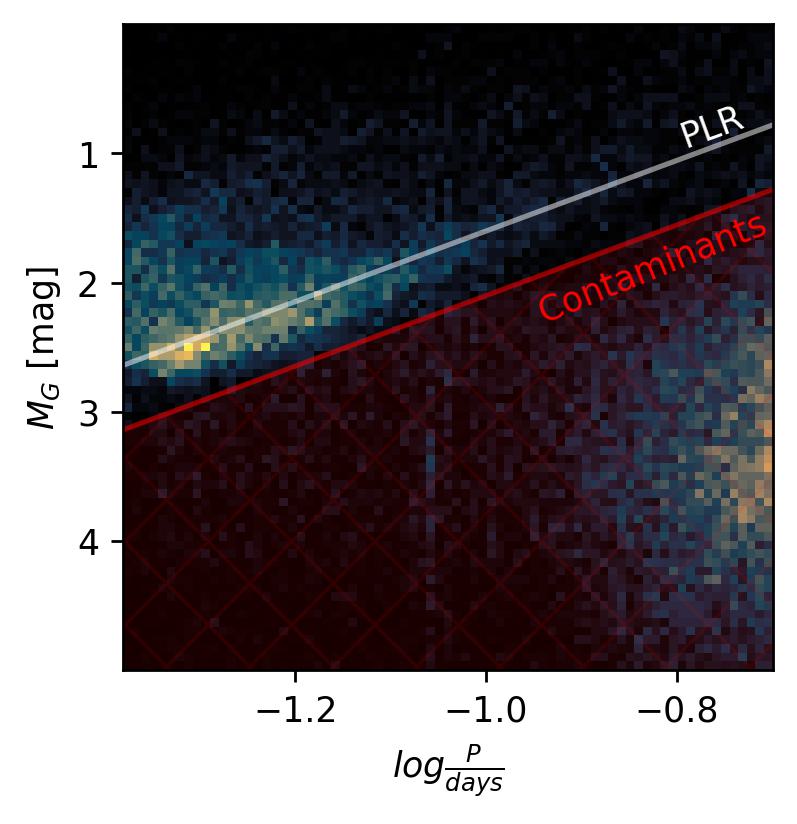}
    \caption{A 2-dimensional histogram showing the density of \textit{Dust Corrected} \textit{Variable Sources} (defined in \S\ref{sec:classification}) as a function of their pulsation periods and absolute $G$-band magnitudes. The line $M_G = -2.734\log P -1.133$, marks the center of the PLR and is plotted as a solid white line. The red hatched region marks where we assume objects are \textit{Non-$\delta$ Scuti Pulsator}s. The lower-right region has a significant over-density of presumed $\gamma$ Dor/hybrid pulsators.}
    \label{fig:plr}
\end{figure}

\begin{figure}[ht!]
    \centering
    \plotone{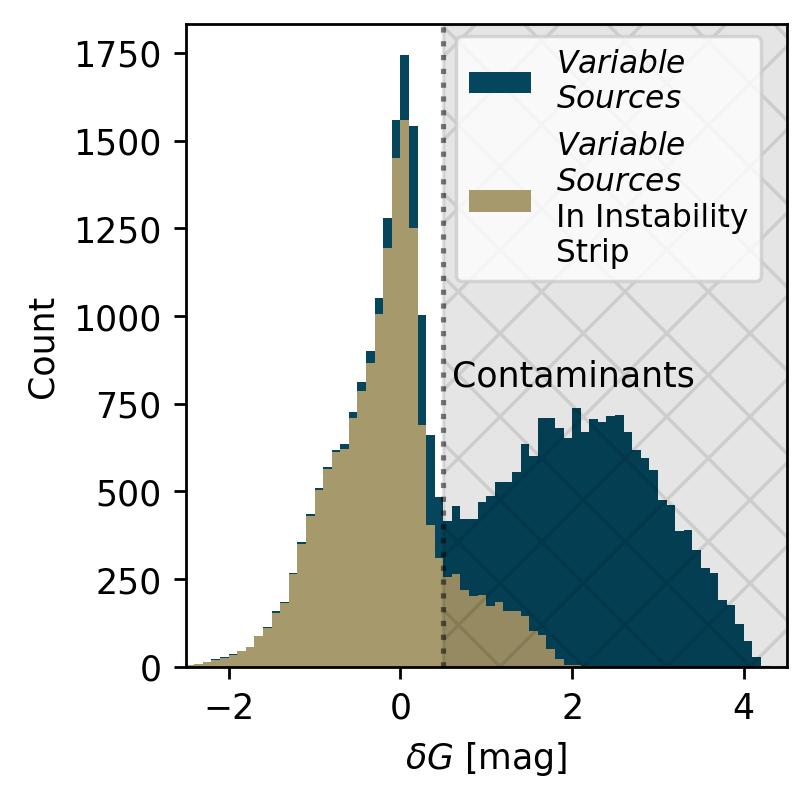}
    \caption{Distributions of differences between objects' observed absolute $G$-band magnitude and the magnitude implied by its pulsation period (i.e. the vertical distance from an object to the white line in Figure \ref{fig:plr}) for all \textit{Dust Corrected} \textit{Variable Sources} in blue and for just \textit{Dust Corrected} \textit{Variable Sources} bluewards of the red edge of} the instability strip in gold. 
    The hatched region marks where we assume objects are \textit{Non-$\delta$ Scuti Pulsators. 
    Because $\delta G$ is in magnitudes, stars on the left-hand side of this plot, with $\delta G < 0$, lie above the PLR marked in Figure \ref{fig:plr}.}
    \label{fig:deltaG}
\end{figure}


To filter out eclipsing binaries we first exploit the shape of their light-curves. Figure \ref{fig:dsctvseblc} shows the first 5 days of the sector 14 QLP light-curve and resulting periodogram of the known EB TYC 2697-130-1 \citep[TIC 14842303; ][]{VillanovaEB}. TYC 2697-130-1 has an orbital period of roughly 1 day, and dips by $\sim10\%$ during each eclipse. The amplitude spectrum shows that the fundamental frequency corresponds to the half period of the orbit, since the primary and secondary eclipses are similar in size. For comparison, the bottom panels of Figure \ref{fig:dsctvseblc} show the light-curve (left) and amplitude spectrum (right) of the sector 16 photometry of $\gamma$ Boo, confirmed as a $\delta$ Scuti star by \citet{BaracPLR}. The periodogram is dominated by a single, strong peak at $\sim13.6$ $d^{-1}$ or $\sim1.77$ hours. Rather than the characteristic shape of an eclipse, the variations in a $\delta$ Scuti star follow a sinusoidal pattern, and therefore do not exhibit the harmonic peaks found in EBs. 

By comparing these two targets we see that, while pulsations cause deviations both above and below the mean flux, eclipses will skew the average flux such that the flux of most measurements will sit above the average, unlike pulsations from a $\delta$ Scuti star. This means that a light-curve with sinusoidal behavior will have roughly half of its flux measurements below the mean, while the flux of an eclipsing binary will be above the mean more often. Thus, we define a metric, $E$, which is the fraction of points in a light-curve below the mean. This metric is similar to skewness, a statistical measurement of the lopsidedness of a distribution \citep[see also ][]{MLClassLC}. 

To calibrate a threshold value for $E$ we compared two smaller samples: a sample of known eclipsing binaries \citep{VillanovaEB} and the sample of $\delta$ Scuti variables from \citet{KeplerDeltaScutiPulsatorFraction}. Figure \ref{fig:dsctebhist} shows the distribution of $E$ for the \emph{TESS} light-curves of stars in each sample. These light-curves are prepared as described in \S\ref{sec:lightcurves}. As expected, the sample of $\delta$ Scuti stars are clustered tightly about $E=0.5$ corresponding to 50\% of flux measurements below each light curve's mean flux. The sample of eclipsing binaries extends from $E=0.5$ to near $0$. Those EBs closer to $E=0.5$ are likely closer binaries which have more sinusoidal light-curves than more detatched binaries which will tend to have lower values of $E$. Only 0.8\% of stars from the sample of $\delta$ Scuti stars fall below $E=0.4$ while nearly $60\%$ of the eclipsing binaries have $E<0.4$. To balance maximizing the number of identified EBs with the number of recovered $\delta$ Scuti, we use $E=0.4$ as a threshold, which yields 22,329 \textit{Eclipsing Binaries} when applied to the \textit{Processed Sample} (defined in \S\ref{sec:samples}).

In fact, this same analysis can find deviations above the mean from pulsations or flares. For example, the high amplitude $\delta$-Scuti stars (HADS) spend less time at their brightest points than their dimmest points, the inverse of what happens in eclipsing systems. Therefore, we can expect HADS to have $E>0.5$. The \emph{TESS} light-curve of the HADS SX Phoenicis (TIC 224285325), for example, yields a value of $E=0.617$. Such stars are quite rare, however, with only a handful found in the \textit{Kepler} field \citep{BalonaHADS}.

\subsubsection{Distinguishing $\delta$ Scuti Stars from Other Pulsators} \label{sec:plr}
$\gamma$ Doradus variables reside redwards of the instability strip, on or near to the main-sequence. They are characterized by their low frequency gravity-mode ($g$-mode) pulsations \citep{hybrids}. $\gamma$ Dor and $\delta$ Scuti variables can be found in overlapping regions of the HRD. Some stars show both the $g$-modes of a $\gamma$ Dor variable and the $p$-mode oscillations of a $\delta$ Scuti variable, known as hybrid pulsators \citep[i.e. ][]{FirstHybrid2002,HybridConfirm2009,hybrids,dsct_hybrids_corot,Balona2018,Antoci2019,Balona2020,Skarka2022}. For a more detailed look at the hybrid population in this analysis see Appendix A.

To help distinguish between $\delta$ Scuti and $\gamma$ Dor pulsators, we use the period-luminosity relation (PLR) of $\delta$ Scuti stars \citep{mcnamara_distance_2000,mcnamara__2007,WesenheitPLR,Ziaali2019,ASASSNdscuti2020,Poro2021,dr3pulsations,Poro2024,BaracPLR,Read2024}. Here, the period ($P$) is the inverse of the frequency of the largest peak in the amplitude-spectrum ($\nu_0$). We observe in Figure \ref{fig:plr} that the majority of \textit{Dust Corrected} \textit{Variable Sources} lie along a diagonal line from roughly $\log P = -1.4$ and $M_G = 2.5$ to $\log P = -0.7$ and $M_G = 1$. 

However, there are also a significant number of stars at $\log P > -0.9$ and $M_G > 2$. These objects are low-luminosity and pulsate near the $5d^{-1}$ limit of our sample . This region of lower luminosity and slower pulsations is characteristic of gravity-mode oscillations \citep{dsct_vs_gdor2002}. Additionally, eclipsing/ellipsoidal/contact binaries, and perhaps even spotted stars, which are not screened out by the $E$-parameter should be similarly clustered alongside the $g$-mode pulsators. Those classes of stars tend to cluster at the low-luminosity end, as low-luminosity stars are most common in general, and at the slower pulsating end, because harmonics of lower frequency pulsations will be strongest at the lowest frequency detected. \citet{prsa2022} searched for eclipsing and ellipsoidal binaries in \textit{TESS} light-curves, and found the shortest period ellipsoidal variables (i.e. morphology parameter near to 1) to have periods generally longer than $\log\frac{P}{days} = -1$.

To measure the fundamental pulsation ridge we assume a relation of the form $G = m ~\log{P} + b$. We then fit a line to the ridge manually as an initial guess. Then, only using stars within a vertical distance of 0.45 magnitudes from that guess, we use \texttt{SciPy}'s \texttt{curve\_fit} routine to perform a least-squares fit. The white line in Figure \ref{fig:plr} marks the resulting fundamental ridge of the PLR:
\begin{equation}
    M_G = (-2.734\pm 0.013) \log P - (1.133\pm 0.015)
\end{equation}

The uncertainties reported above are formal errors on the fit, and do not account for systematic errors. The main systematic error, imposed by the 30 minute cadence light-curves we use, is the effect of Nyquist aliasing. 
\citet{Read2024} shows that nearly half of the $\delta$-Scuti stars that they analyze pulsate faster than 24 $d^{-1}$. In our analysis such stars will be reflected horizontally across the left side of Figure \ref{fig:plr}. Such smearing of the PLR of the fundamental radial pulsation mode introduces additional uncertainty to the above fit which is not accounted for.

However, there are clearly more stars above this ridge than below, particularly in the region $\log P < -1.2$, near to the Nyquist frequency. Many $\delta$ Scuti stars pulsate in the first or higher overtone rather than the fundamental mode. This manifests as a second ridge above the fundamental ridge \citep[an example can be found in ][]{BaracPLR}. By analyzing the distribution of $\delta G$---the vertical distance from the PLR---in Figure \ref{fig:deltaG}, we identify that the peak at $\delta G=0~mag$ corresponds to the fundamental ridge while the overtone ridge manifests as an over density near $\delta G=-1~mag$, and the likely \textit{Non-$\delta$ Scuti Pulsators} manifest as a wide distribution of objects in the hatched region of Figure \ref{fig:deltaG} ($\delta G>0.5~mag$).

\begin{figure}
    \centering
    \plotone{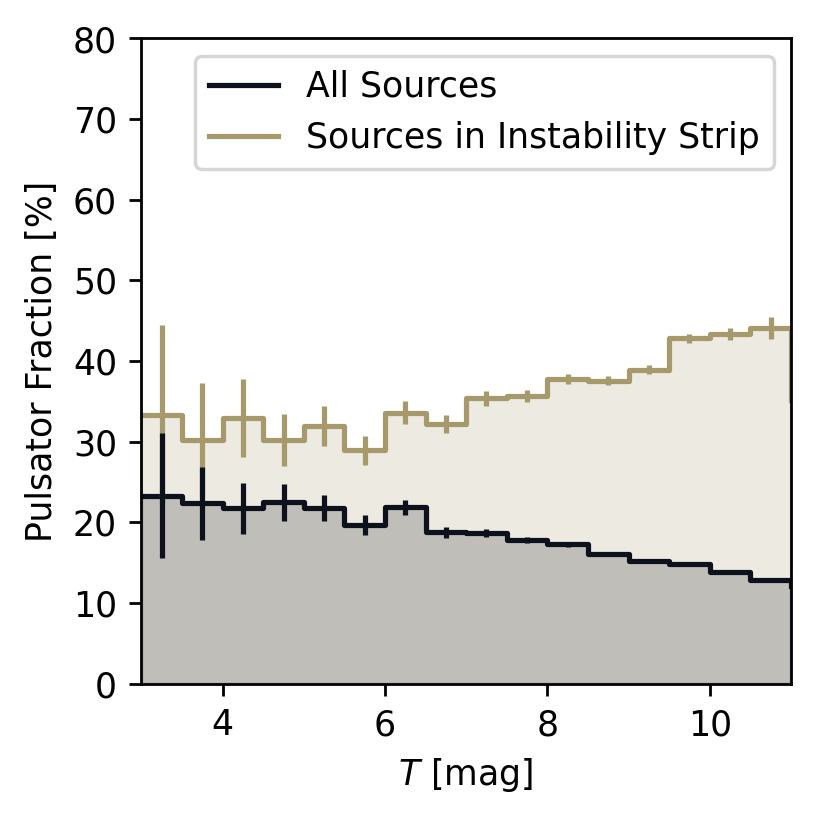}
    \caption{A plot showing the mean pulsator fraction across all colors as a function of apparent $T$-magnitude for all sources in black, and for only sources bluewards of the red edge of} the instability strip in gold. Vertical lines indicate uncertainties calculated using binomial statistics, $\sigma=\sqrt{\frac{p(100-p)}{N}}$, where $p$ is the percentage of sources from the \textit{Processed Sample} in each bin which are also \textit{Variable Sources} and $N$ is the total number of sources from the \textit{Processed Sample} in each bin (defined in \S\ref{sec:samples}).
    \label{fig:tpulse}
\end{figure}

If we assume that only stars with $\delta G < 0.5~mag$ are \textit{$\delta$ Scutis}, then only 15,918 out of 34,061 stars, or $\sim$47\% of our \textit{Dust Corrected} \textit{Variable Sources} (that is members of both the \textit{Dust Corrected}, and \textit{Variable Sources} samples) are true $\delta$ Scuti pulsators. However, this sample includes stars outside of the instability strip, where $\delta$ Scuti type pulsations are not expected. When considering only \textit{Variable Sources} bluewards of the red edge of the instability strip (to be defined in \S\ref{sec:instabilitystrip}), 14,307 out of 16,541, or over 86\% of \textit{Dust Corrected} \textit{Variable Sources} are \textit{$\delta$ Scutis}, as shown in Figure \ref{fig:deltaG}. Meanwhile the fraction of \textit{Non-$\delta$ Scuti Pulsators} are drastically reduced, showing that the majority of \textit{Non-$\delta$ Scuti Pulsators} are redwards of the classical instability strip as expected.

\begin{center}
\begin{deluxetable*}{l|ccccccccc}  
    \tablecolumns{12}
    \setlength{\tabcolsep}{2.9pt}
    \tablecaption{Examples of values derived in this work and tabulated in this catalog for a random selection of \textit{Variable Sources}}
    \tablehead{\colhead{\emph{TESS} ID} &  
    \colhead{Variable\tablenotemark{a}} &  
    \colhead{Best Sector\tablenotemark{b}} &
    \colhead{EB\tablenotemark{c}} &  
    \colhead{$\delta$ Scuti\tablenotemark{d}} &  
    \colhead{$\nu_0$\tablenotemark{e}} &  
    \colhead{$A_0$\tablenotemark{f}} &  
    \colhead{$N_{harmonics}$\tablenotemark{g}} & 
    \colhead{$E(G_{BP}-G_{RP})$\tablenotemark{h}} &  
    \colhead{$A_G$\tablenotemark{i}} \\ 
     & & & & & \colhead{[$d^{-1}$]} & \colhead{[$ppm$]} & & \colhead{[$mag$]} & \colhead{[$mag$]}}
    \startdata
        16780066 & 1 & 21  & False & 0 & 18.74 & 899.6  & 0   & --- & --- \\
        16781787 & 1 & 15  & False & 1 & 6.001 & 787.7  & --- & 0.0399 & 0.1018 \\
        16790069 & 1 & 12  & False & 0 & 6.928 & 3964   & --- & --- & --- \\
        16790980 & 1 & 12  & False & 1 & 11.24 & 1619   & 0   & 0.0854  & 0.2180 \\
        16794208 & 1 & 12  & False & 0 & 8.612 & 3166   & 0   & --- & --- \\
        16801815 & 1 & 12  & False & 0 & 6.928 & 4656   & --- & 0.0716 & 0.1827  \\
        16805775 & 1 & 21  & True  & 0 & 5.682 & 266.6  & --- & --- & --- \\
        16807906 & 1 & 21  & False & 0 & 17.53 & 2456   & 0   & --- & --- \\
        16810156 & 1 & 17  & False & 1 & 9.505 & 6812   & 0   & 0.0447  & 0.1142  \\
        16810165 & 1 & 17  & False & 0 & 9.509 & 279.0  & 0   & 0.0516 & 0.1316 \\
        16834725 & 1 & 14  & False & 0 & 17.74 & 339.3  & --- & --- & --- \\
        16836537 & 1 & 14  & False & 0 & 6.455 & 484.6  & --- & --- & --- \\
        16836798 & 1 & 14  & False & 0 & 9.151 & 342.4  & --- & --- & --- \\
        16843848 & 1 & 15  & False & 0 & 9.713 & 519.7  & 0   & --- & --- \\
        16844778 & 1 & 14  & False & 0 & 18.25 & 490.6  & --- & --- & --- \\
        16845152 & 1 & 14  & False & 0 & 5.294 & 311.5  & --- & 0.0267 & 0.0681 \\
        16880980 & 1 & 17  & False & 0 & 6.200 & 11950  & 0   & --- & --- \\
    \enddata
    \label{table}
    \tablenotetext{a}{denotes whether the object has at least one statistically significant peak in its periodogram between 5 and 24 $d^{-1}$ (\S\ref{sec:classification}).}
    \tablenotetext{b}{the \textit{TESS} sector whose amplitude spectrum has the largest amplitude at $\nu_0$ across all available sectors.}
    \tablenotetext{c}{denotes whether the object is classified as an \textit{Eclipsing Binary} with the $E$-metric described in \S\ref{sec:class_eb}.}
    \tablenotetext{d}{a classification of variable sources as true \textit{$\delta$ Scutis} (1) or \textit{Non-$\delta$ Scuti Pulsator}s (0) based on vertical distance from period-luminosity relation as described in \S\ref{sec:plr}. Dashes indicate that classification could not be performed, either because we could not derive $A_G$ or because fractional distance uncertainties exceeded 20\%.}
    \tablenotetext{e}{the frequency of the highest amplitude peak between 5 $d^{-1}$ and 24 $d^{-1}$. }
    \tablenotetext{f}{the amplitude corresponding to $\nu_0$, or the height of the peak associated with $\nu_0$.}
    \tablenotetext{g}{the number of harmonic peaks detected corresponding to integer-multiples of $\nu_0$. Dashes indicate that there were not any additional, statistically significant, peaks which could be harmonics. Zero indicates that there were other statistically significant peaks, but they were not harmonics of $\nu_0$.}
    \tablenotetext{h}{reddening in \textit{Gaia} color from interstellar dust, calculated from the \citet{LiekeDustMap} dust map. \textbf{Dashes indicate dust-estimation failed, most commonly due to the object falling outside of the dust map.}}
    \tablenotetext{i}{extinction in $G$-band due to interstellar dust, calculated from the \citet{LiekeDustMap} dust map. \textbf{Dashes indicate dust-estimation failed, most commonly due to the object falling outside of the dust map.}}
\end{deluxetable*}
\end{center}
\newpage
\subsection{Summary of Stars Analyzed} \label{sec:samples}
We summarize our sample selections as follows:

\begin{itemize}
    \item \textit{Processed Sample}\footnote{In the interest of clarity, when we are refering to one of the following samples of stars we will capitalize and italicize it in this text. For example, while we may refer to the broader population of $\delta$ Scuti variable stars, \textit{$\delta$ Scutis} refer to the specific group of stars defined in this paper as $\delta$ Scuti variables.}---\S\ref{sec:sampleselction}---754,909---All stars from the \textit{Full Sample} with $T<11.25$ and for which at least one \emph{TESS} sector of data between sectors 1 and 26 is available.
    \item \textit{Variable Sources}---\S\ref{sec:classification}---103,810---All stars from the \textit{Processed Sample} which have at least one statistically significant peak in its Lomb-Scargle periodogram between 5 and 24 cycles per day. \textbf{Of these sources, 34,061 are also \textit{Dust Corrected} as described below.}
    \item \textit{Non-variable Sources}---\S\ref{sec:classification}---651,099---All stars from the \textit{Processed Sample} which did not have any statistically significant peaks in its Lomb-Scargle periodogram between 5 and 24 cycles per day.
    \item \textit{Dust Corrected}---\S\ref{sec:dust}---439,861---All stars from the \textit{Processed Sample} for which we could correct for the effects of interstellar dust using the \citet{LiekeDustMap} dust-map.
    \item \textit{Eclipsing Binaries}---\S\ref{sec:class_eb}---22,229---All stars from the \textit{Processed Sample} which have $E<0.4$.
    \item \textit{$\delta$ Scutis}---\S\ref{sec:plr}---\textbf{15,918}---All \textit{Variable Sources} which are \textit{Dust Corrected} and which do not lie below the period-luminosity relation.
    \item \textit{Non-$\delta$ Scuti Pulsators}---\S\ref{sec:plr}---20,488---All \textit{Variable Sources} which are \textit{Dust Corrected} and which lie below the period-luminosity relation.
    \item \textit{Non-variable Instability Strip Stars}---\S\ref{sec:instabilitystrip}---28,169---All \textit{Non-variable Sources} which reside bluewards of the red edge of the instability strip.
\end{itemize}

Table \ref{table} lists the quantities derived in this work. For each TIC ID analyzed we have additionally compiled information from 2 sources: the \emph{TESS} Input Catalog \citep[TIC; ][]{TIC} and the \textit{Gaia} DR3 source catalog \citep{smith_Gaia_2012}. 

\begin{figure*}[ht!]
    \centering
    \plotone{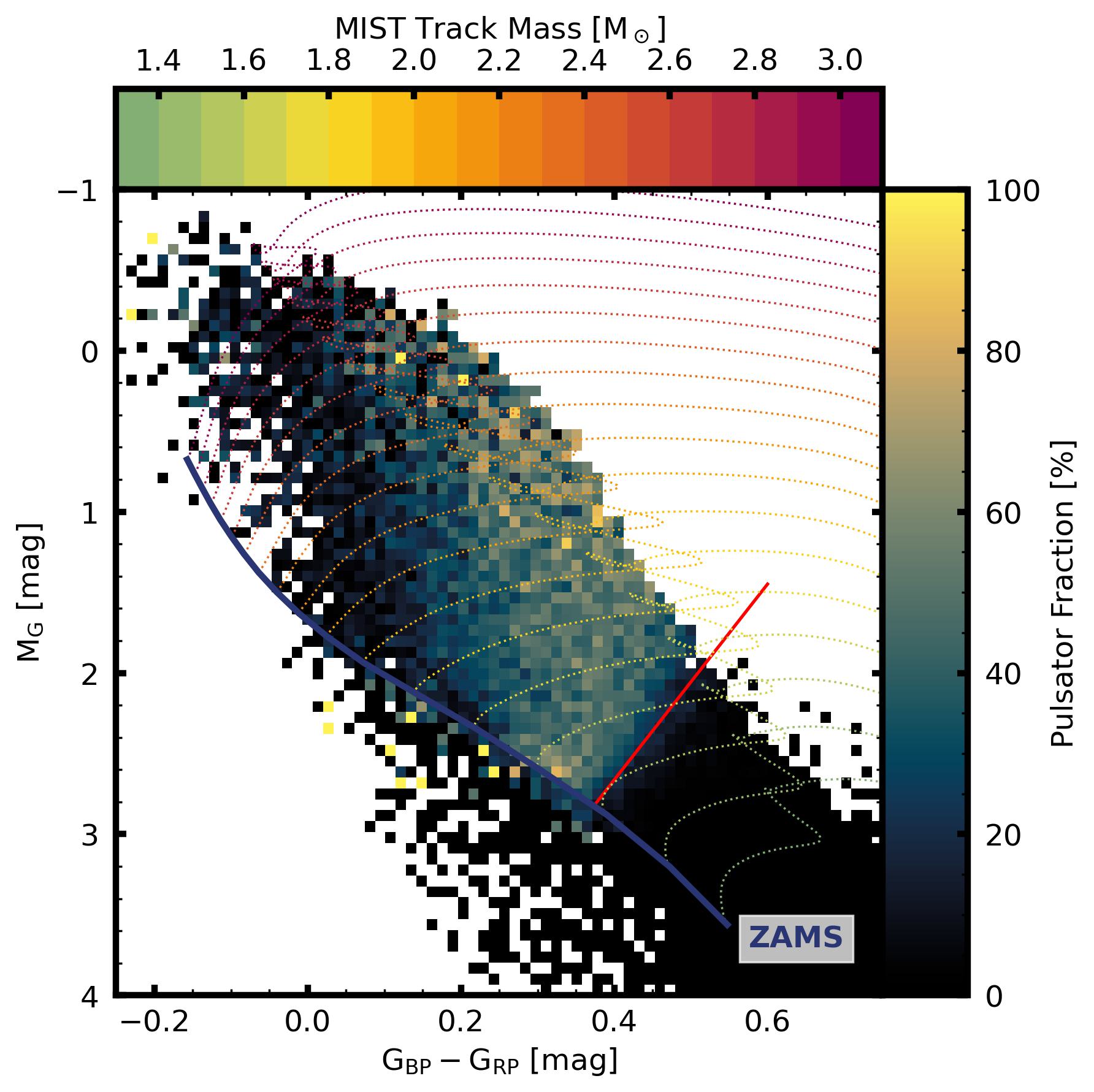}
    \caption{A color-magnitude diagram in which each bin is colored to represent the percentage of stars in that bin which we classify as \textit{$\delta$ Scutis} via PLR classification (\S\ref{sec:plr}).
    Only bins with at least 5 stars are colored. 
    We additionally over-plot the red edge (Eq. \ref{eq:red}) of the instability strip, chosen to align with the 20\% contour. 
    Solar metalicity, rotating MIST models of stars between 1.3 and 3.1 $M_\sun$ are plotted as dotted lines, with their bases connected to highlight the zero age main-sequence.
    }
    \label{fig:pulse_frac}
\end{figure*}

\begin{figure}
    \centering
    \plotone{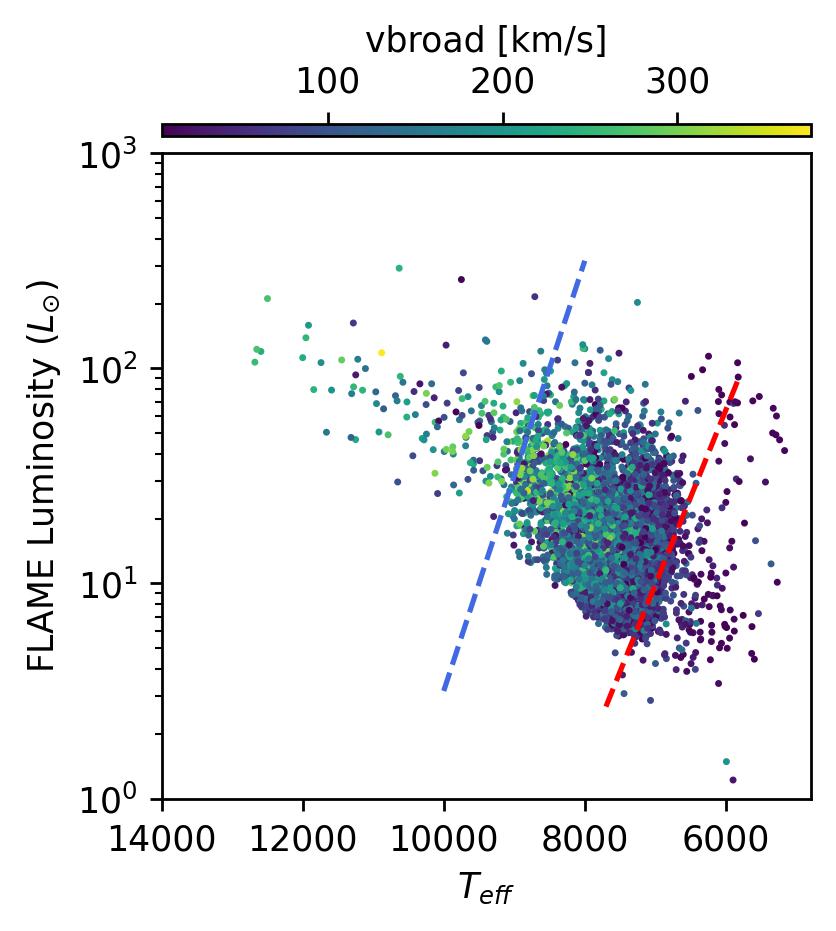}
    \caption{A Hertzsprung-Russell diagram of $\delta$ Scuti variables, colored by the Gaia line broadening measure \texttt{vbroad}. Luminosities are from FLAME \citep{FLAME}, while effective temperatures are from two sources. For the hotter stars ($T_{\mathrm{eff}} > 9000~K$) we adopt spectroscopic \textit{Gaia} ESPHS measurements, while for stars cooler than $9000\,$K we adopt GSPPHOT measurements \citep{GaiaESPHS} because the ESPHS coverage does not include the entire instability strip. Dashed lines show the instability strip reported in \citet{KeplerDeltaScutiPulsatorFraction}.}
    \label{fig:hrd}
\end{figure}

Figure \ref{fig:tpulse} shows that in general our variability fraction declines as $T$-magnitude increases. 
This demonstrates that the completeness of our method is limited by the brightness of the sample. 
However, when we limit our analysis to stars bluewards of the red edge of the instability strip we see the opposite trend. 
This can be explained as an effect of missing quickly pulsating, blue, $\delta$ Scuti stars. 
Those blue stars, in a volume limited sample, will tend to be apparently brighter than red stars which are more likely to have pulsations detected.

\section{An Empirical Instability Strip} \label{sec:instabilitystrip}

Figure \ref{fig:pulse_frac} shows pulsator fraction (the percentage of stars in each bin which are \textit{$\delta$ Scuti} pulsators) as a function of location on the CMD. 
Following \citet{KeplerDeltaScutiPulsatorFraction}, we detect the instability strip as a pronounced ridge in the pulsator fraction plot. 
The pulsator fraction rises to nearly 70\% at its two highest points near to $G_{BP}-G_{RP} \simeq 0.3$, $M_G = 2.5$ and $0 < M_G < 1.5$. Higher pulsator fractions among more luminous stars are also seen in \citet{KeplerDeltaScutiPulsatorFraction}, and is therefore likely astrophysical in nature. We hypothesize that this is due to increase in pulsation amplitude with luminosity (similar to convection driven oscillators).

Although direct comparison is not possible because \citet{KeplerDeltaScutiPulsatorFraction} used effective temperatures and luminosities rather than color and magnitude, the similarity in peak pulsator fraction suggests that Kepler's selection function did not bias the population statistics derived from $\delta$ Scuti stars within the \textit{Kepler} field.
Further, despite the difficulties in converting to physical measurements, Figure \ref{fig:hrd} shows that for a subset of \textit{$\delta$ Scutis} the instability strip reported in \citet{KeplerDeltaScutiPulsatorFraction} fits our sample well.

Following \citet{KeplerDeltaScutiPulsatorFraction}, we attempt to create boundaries to this instability strip by delineating the region where pulsator fraction rises to $\sim$20\%. We do this by drawing 20\% contour lines over Figure \ref{fig:pulse_frac}, and extracting the vertices of that contour line on the red edge. We then use SciPy's \texttt{curve\_fit} routine to fit a straight line to this edge, leading to the following line:
\begin{equation} \label{eq:red}
    M_G = (5.082 \pm 0.012) - (6.054 \pm 0.063) (G_{BP}-G_{RP}).
\end{equation}

We attempted the same process on the blue side, however the sparsity of stars prevented a reasonable fit. This is likely due to a combination of this method missing fast-pulsating $\delta$ Scuti stars which are disproportionately hotter, higher mass stars, and blending of $\delta$ Scuti stars with hotter OB stars. Hot main-sequence stars are believed to be the most rapidly pulsating $\delta$ Scuti stars, which can reach pulsation frequencies more than double our 24 $d^{-1}$ Nyquist limit \citep{TESSPleadies}. Due to our inability to distinguish fast pulsators from non-pulsators, we do not define a blue edge. Higher cadence data could capture these rapidly pulsating $\delta$ Scuti stars, and allow for such a fit.

\begin{figure}
    \centering
    \plotone{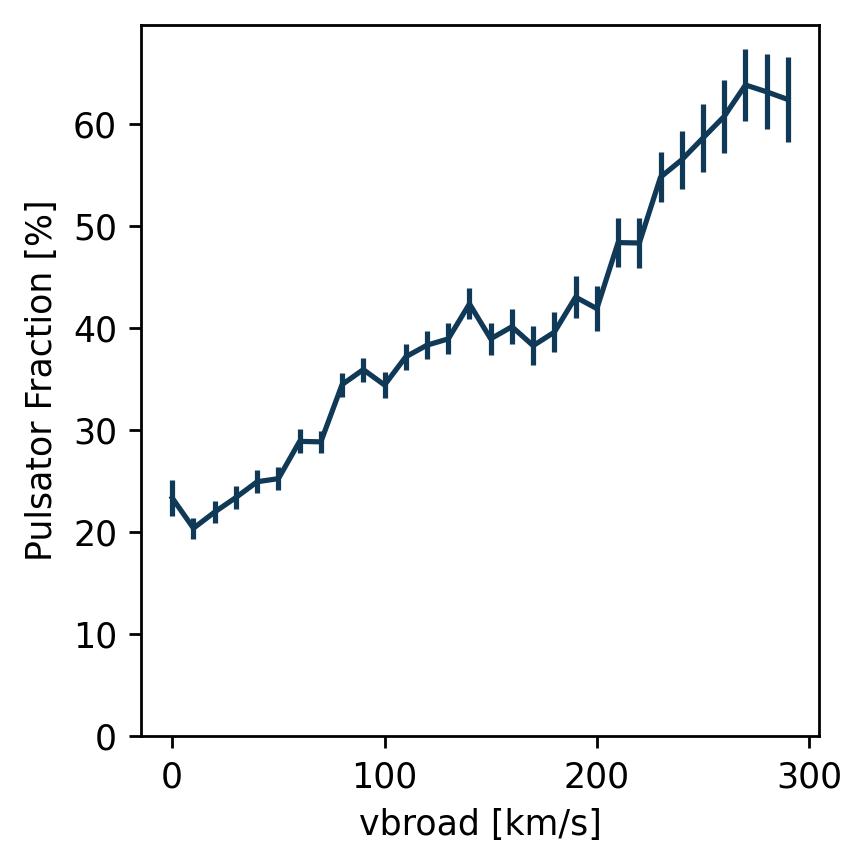}
    \caption{Pulsator fraction (the percentage of \textit{Dust Corrected} stars bluewards of the red edge of the instability strip in each bin classified as \textit{$\delta$ Scutis}) as a function of \texttt{vbroad}. Vertical lines indicate uncertainties calculated using binomial statistics, $\sigma=\sqrt{\frac{p(100-p)}{N}}$, where $p$ is the percentage of sources from the \textit{Processed Sample} in each bin which are also \textit{Variable Sources} and $N$ is the total number of sources from the \textit{Processed Sample} in each bin.}
    \label{fig:vbpf}
\end{figure}

\begin{figure*}
    \centering
    \plottwo{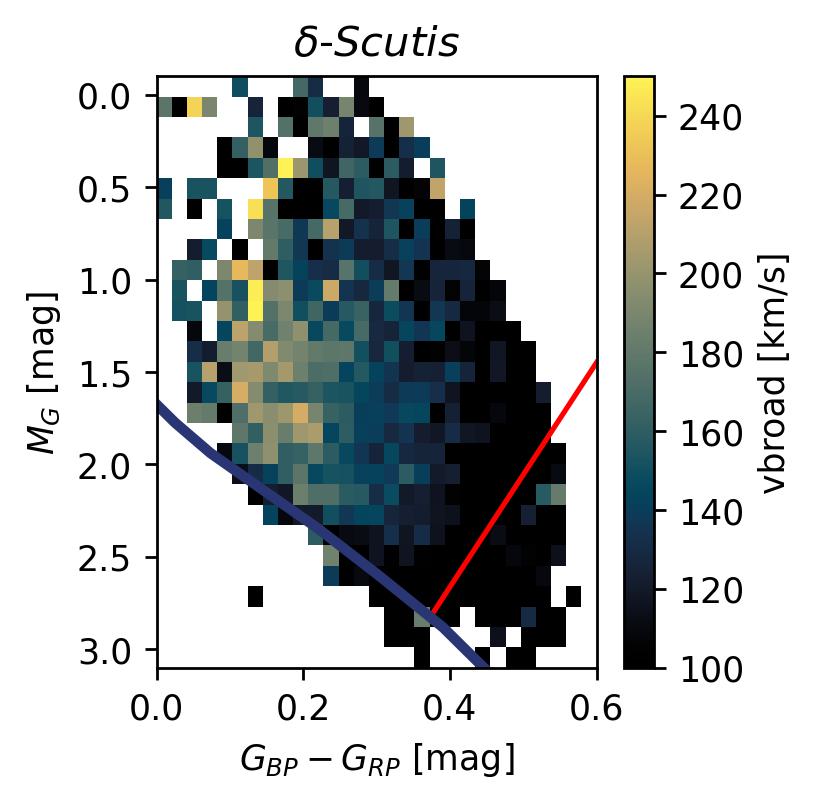}{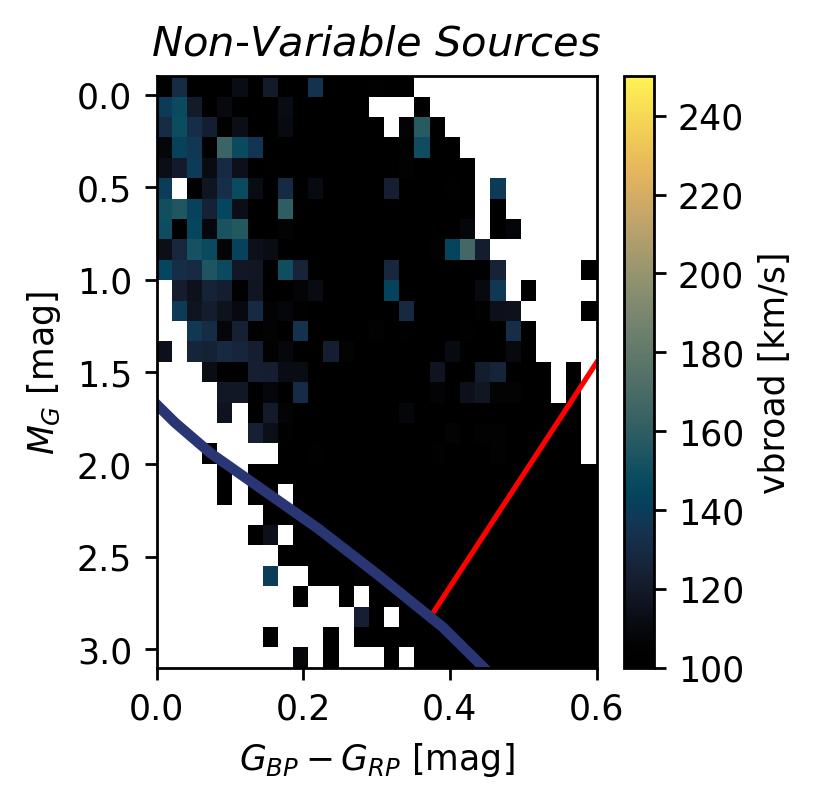}
    \caption{A CMD colored by the mean \texttt{vbroad} values of \textit{$\delta$ Scuti} variables (left) and \textit{Non-variable}s (right) in each bin. Only bins with at least 10 objects are shown. Note that any bin with a value below 100 km/s is colored black, meaning that \textit{Non-variable}s---over most of the cmd---have values of \texttt{vbroad} below 100 km/s. \textit{$\delta$ Scuti}s, however, show \texttt{vbroad} values over 100 km/s over most of the instability strip. We additionally over-plot lines representing the red edge of the instability strip.}
    \label{fig:vbdiff}
\end{figure*}

In Figure \ref{fig:pulse_frac}, we over-plot the red edge of the instability strip, chosen to align with the 20\% contours, where inside this empirical instability strip the pulsator fraction is at least $\sim$ 20\%. We additionally plot a series of MIST evolutionary tracks for solar metallicity, rotating ($\frac{v}{v_{crit}}=0.4$) stars with masses between 1.3 and 3.1 $M_\sun$. The ZAMS, which connects the bases of these evolutionary tracks, has a considerable number of stars below it. The variable stars below the main sequence are likely variable sub-dwarfs or delta Scuti stars with erroneous reddening corrections. The stars, however, make up a small fraction of the overall sample (each bin below the ZAMS contains the minimum of 5 stars whereas above the ZAMS each bin contains dozens to hundreds of stars).

\section{A Correlation Between Pulsator Fraction and Rotation} \label{sec:vbroad}
Our large and homogeneous catalog allows us to investigate the occurrence of $\delta$ Scuti pulsation as a function of other physical parameters such as stellar rotation. Figure \ref{fig:vbpf} shows a striking correlation of pulsator fraction with \texttt{vbroad}, this parameter is a measure of spectral line broadening from \textit{Gaia}'s RVS spectrograph \citep{RVSDR3,GaiaRVS}. \texttt{vbroad} is a measurement of all factors which might contribute to spectral line broadening (including $v\sin i$, mirco-turbulence, and macro-turbulence). \citet{GaiaRVS} shows that for most ranges of temperature and magnitude, \texttt{vbroad} is nearly equivalent to independently measured values of $v\sin i$ up to approximately 100-200 $km~s^{-1}$, depending on the temperature and brightness of the star. We have additionally attempted the same analysis with other sources, such as \textit{Gaia}'s \texttt{vsini}$_{esphs}$ parameter which attempts to disentangle $v\sin{i}$ from other spectral line broadening phenomena \citep{GaiaESPHS}, as well as $v\sin{i}$ measurements from the Apache Point Observatory Galactic Evolution Experiment \citep[APOGEE; ][]{APOGEE}. Both of these parameters lead to a similar observed correlation.

Since stellar $v\sin i$ is also a function of color (see Figures \ref{fig:hrd} \& \ref{fig:vbdiff}), where hot stars generally rotate more rapidly than cool stars \citep{KraftBreak}, one simple explanation is that Figure \ref{fig:vbpf} is showing that \textit{$\delta$ Scuti} stars are on average bluer than \textit{Non-variable Sources}. However, as is shown in Figure \ref{fig:vbdiff}, \textit{$\delta$ Scutis} have larger \texttt{vbroad} than \textit{Non-variable Sources}, even over small regions of the CMD. The left plot of Figure \ref{fig:vbdiff} represents mean \texttt{vbroad} over the CMD for \textit{$\delta$ Scuti} stars, showing that over the majority of the instability strip stars rotate at velocities between 100 and 250 km/s. The \textit{Non-variable Sources} are shown on the right. Over the majority of the instability strip, \textit{Non-variable Sources} rotate at velocities below 100 km/s.

One possible explanation for this effect is rotational mixing \citep{rotation_winds_owecki_1996,rotmixinghotstars,rotmixinghuang}. Since the $\kappa$-mechanism relies on the partial ionization of helium in a specific layer of a star's atmosphere, a star requires helium in that layer for classical pulsations to occur. As stars age, if they are not steadily mixed, they become chemically stratified, as heavier elements gravitationally settle to lower layers within the star. This is in agreement with findings of low pulsation fractions among the slow-rotating, metallic-lined A-stars \citep{AmPulse1970,dsctRotation1974,AmPulse1977,dsctRot2015}. Should helium gravitationally settle below the partial ionization layer in a $\delta$ Scuti variable, the driving of those pulsations will weaken. As described by \citet{Murphychemnorm}, non-pulsators in the instability strip may be those that are magnetically active and slowly rotating. Our catalog yields a sample of bright, \textit{Non-variable Instability Strip Stars}, which are ideal targets for follow-up with high resolution spectroscopy to confirm this hypothesis.

The small number of rapidly rotating \textit{Non-variable Sources} on the blue side of the right panel of Figure \ref{fig:vbdiff} may also suggest that the analysis in this work fails in identifying fast pulsating, blue, $\delta$ Scuti stars. Near to the red edge, where pulsator fraction drops towards zero the \textit{Non-$\delta$ Scuti Pulsators} rotate below 100 km/s. On the blue side ($0< G_{BP}-G_{RP} < 0.2$), however, the \textit{Non-variable sources} show significantly higher values of \texttt{vbroad}, between 100 and 150 km/s. If pulsation does correlate with rotation velocity a targeted study of quickly rotating stars in that region of the CMD should reveal $\delta$ Scuti stars which are missed in this analysis.

\section{Conclusion} \label{sec:con}
By analyzing nearly one million 30-minute cadence \emph{TESS} QLP light-curves with $T<11.25$ \citep{QLP1,QLP2,QLPextended} we have identified variability in 103810 sources and confirmed \textbf{15,918} \textit{$\delta$ Scuti} variables. This is an order of magnitude leap in the search for classical pulsators compared to the \textit{Kepler} field and utilizes data with an extremely simple selection function. Our main conclusions are as follows:
\begin{itemize}
    \item We measure a period-luminosity relation ($M_G = (-2.734\pm 0.013) \log_{10} P - (1.133\pm 0.015)$ where $P$ is the pulsation period measured in days) and identify contaminating stars which are presumably either EBs, hybrid $p/g$-mode pulsators, or $g$-mode pulsators. We discuss how limitations of 30-minute cadence data affect this relation, but demonstrate by manually classifying 200 stars in Appendix A that it is sufficient for classification ($<7\%$ false negative rate and $<2\%$ false positive rate).

    \item After identifying \textit{$\delta$ Scuti} pulsators, we calculate pulsator fraction across the color-magnitude diagram (CMD) and produce a red boundary for an empirical instability strip in observed \textit{Gaia} parameters---$G_{BP}-G_{RP}$ and $M_G$ (equation \ref{eq:red}). Consistent with previous investigations we find that, even in the center of the instability strip over 20\% of sources show no pulsations.
    
    \item We show that \textit{Gaia}'s \texttt{vbroad} parameter---a measure of spectral line broadening---is systematically larger for \textit{$\delta$ Scuti} sources than for their non-pulsating cousins. This pattern holds even when controlling for location in the CMD. This correlation confirms that rotation plays a crucial role in sustaining classical pulsations. We hypothesize that more slowly rotating A-F stars become more chemically stratified, allowing helium to gravitationally settle below the partial ionization layer where the $\kappa$-mechanism drives classical pulsations.
\end{itemize}

This work naturally suggests a few lines of inquiry to follow in the future. For one, using higher cadence data will allow for more reliable identification of rapidly pulsating stars, which will allow for a more reliable PLR, as well as drawing the blue edge of the instability strip. Further, the procedures outlined in this work could be replicated for other types of pulsators. For example, $\gamma$ Doradus pulsators reside within similar regions of the CMD. $\gamma$ Doradus variables pulsate in non-radial $g$-modes and simply extending the range of frequencies analyzed would likely find a significant number of $\gamma$ Doradus variables in the region red-ward of the instability strip. Similarly, higher-cadence data, such as the 600s cadence data available in later \emph{TESS} cycles, could capture higher-frequency pulsators, such as young $\delta$ Scuti stars \citep{FastPulsatingDeltaScutiNature}. Finally, an analysis including pixel-level data analysis \citep[e.g. ][]{TESSLocalize} would be helpful to disentangle different sources of variability in crowded fields. For example, in Table \ref{table}, TIC 16810165 and TIC 16810156 have the same dominant frequency, and are next to each other on the sky. This pixel level data would be useful in determining which object belongs to the 9.5 $d^{-1}$ signal.

\begin{acknowledgments}
We thank the referee for their timely and helpful comments, which inspired the below appendix, and pointed out the potential for the $E$ parameter described in \S\ref{sec:class_eb} to detect HADS such as SX Phoenicis. 

K.G.\ and D.H.\ acknowledge support from the the National Aeronautics and Space Administration through the \emph{TESS} General Investigator Program (80NSSC21K0784). M.H. acknowledges support from NASA through the NASA Hubble Fellowship grant HST-HF2-51459.001 awarded by the Space Telescope
Science Institute, which is operated by the Association of Universities for Research in Astronomy, Incorporated, under NASA contract NAS5-26555.
D.H.\ also acknowledges support from the Alfred P. Sloan Foundation and the Australian Research Council (FT200100871). 
D.R.H.\ acknowledges support from the National Science Foundation (AST-2009828).
T.R.B.\ and S.J.M.\ gratefully acknowledge support from the Australian Research Council through Future Fellowship FT210100485 and Laureate Fellowship FL220100117.

Funding for the \emph{TESS} mission is provided by NASA's Science Mission directorate. This paper includes data collected by the \emph{TESS} mission, which are publicly available from the Mikulski Archive for Space Telescopes (MAST).

This work has made use of data from the European Space Agency (ESA) mission
{\it Gaia} (\url{https://www.cosmos.esa.int/gaia}), processed by the {\it Gaia}
Data Processing and Analysis Consortium (DPAC,
\url{https://www.cosmos.esa.int/web/gaia/dpac/consortium}). Funding for the DPAC
has been provided by national institutions, in particular the institutions
participating in the {\it Gaia} Multilateral Agreement.

\end{acknowledgments}

\software{This research made use of \texttt{NumPy} version 1.24.3 \citep{numpy}, 
\texttt{pandas} version 2.1.4 \citep{pandas,pandas-v2.1.4}, 
\texttt{AstroPy} version 5.3.4 \citep{astropy2013,astropy2018,Astropyv5}, 
\texttt{SciPy} version 1.11.4 \citep{scipy}. 
All plots were made using \texttt{matplotlib} version 3.7.2 \citep{matplotlib, matplotlib-v3.7.2}, 
Figure \ref{fig:hrd} makes use of \texttt{matplotlib}'s \texttt{viridis}  colormap,
and the remaining plots use the \texttt{eclipse} colormap from \texttt{CMasher} \citep{cmasher}.}

\bibliographystyle{aasjournal}
\bibliography{main}

\appendix
\section{Manual Classification}
In order to assess the accuracy of our automated methods, we take a closer look at a subsample of 200 objects. The stars were selected as to sample stars across our period-luminosity diagram (Figure \ref{fig:plr}). Using \texttt{NumPy}'s \texttt{choice} method, we sample 200 bins, without replacement, using the number of objects in each bins as weights. This method chooses 200 different locations in Figure \ref{fig:plr}, with populated areas of the diagram most likely to be chosen. We then randomly choose one star from each bin to analyze.

In order to classify each star we produce a series of diagnostic plots. Each set of diagnostic plots include the following:
\begin{itemize}
    \item Light-curve: the light-curve of each available \emph{TESS} sector. The best sector (the sector whose amplitude spectrum contains the highest peak) is plotted in gold while all others are plotted in blue.
    \item Amplitude Spectrum: the amplitude spectrum of the best sector. In addition to the frequency space analyzed between 5 and 24 $d^{-1}$, we plot as low as 1 $d^{-1}$ in order to check for lower frequency $g$-modes which would indicate an object is a hybrid-pulsator. $\nu_0$ is marked by a black line.
    \item Color-Magnitude Diagram: we reproduce Figure \ref{fig:pulse_frac}, and show the location of the object in relation to the instability strip as a pink star.
    \item Period-Luminosity Diagram: we reproduce Figure \ref{fig:plr}, and show the location of the object in relation to the PLR.
\end{itemize}

We then classify each object with three questions:
\begin{enumerate}
    \item is the automated classification correct? These are sorted into true-positive (TP), true-negative (TN), false-positive (FP), false-negative (FN).
    \item Does this object appear to be a hybrid pulsator? This is judged by the presence of both low frequency and high frequency peaks in the amplitude spectrum. High frequency peaks should be sufficiently high so that they would be on or above the PLR plotted in Figure \ref{fig:plr}. Low frequency peaks should be sufficiently low so that they would be in the red hatched region in Figure \ref{fig:plr}.
    \item Does this object appear to be an eclipsing binary (EB) based on the shape of the light-curve and amplitude spectrum?
\end{enumerate}

\begin{figure*}[ht!]
    \plotone{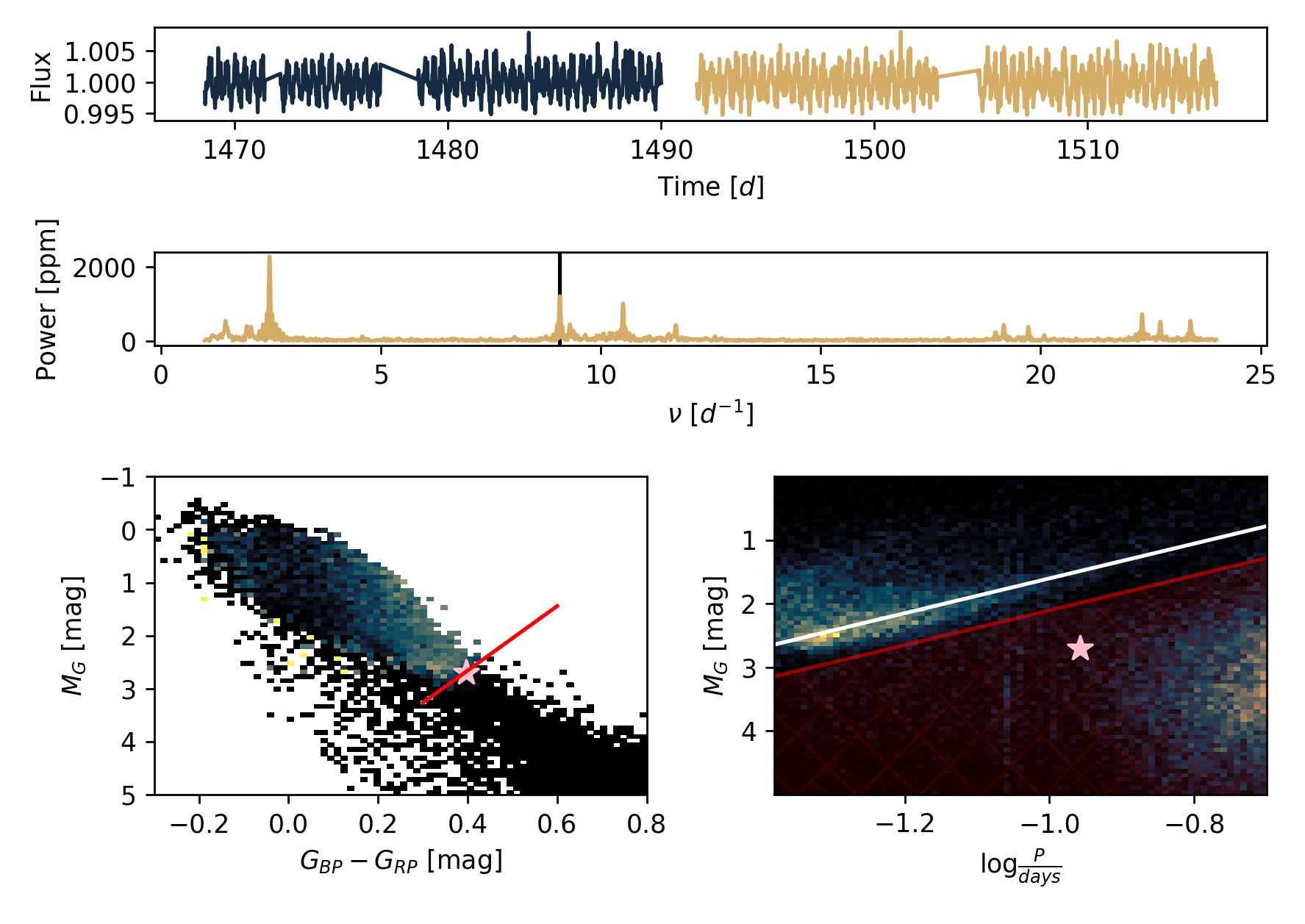} 
    \caption{Diagnostic plot of  TIC 172309348. \textit{Top:} QLP light curves of \textit{TESS} sectors 6 in blue and 7 in gold. \textit{Middle:} Amplitude spectrum of the sector 7 light curve for $1~d^{-1} < \nu < 24~d^{-1}$. \textit{Bottom left:} A replication of Figure \ref{fig:pulse_frac} with the location of TIC 172309348 marked with a pink star. \textit{Bottom right:} A replication of Figure \ref{fig:plr} with the location of TIC 172309348 marked with a pink star.}
    \label{fig:diag_example}
\end{figure*}

We show an example diagnostic plot in Figure \ref{fig:diag_example} of TIC 172309348. The amplitude spectrum in the middle panel is very interesting, with statistically significant peaks across the spectrum. The largest peak in the spectrum is around $\nu=2~d^{-1}$, with the next largest being a group of peaks around $\nu=10~d^{-1}$, finally there are a group of smaller, but significant, peaks between $20~d^{-1}<\nu<24~d^{-1}$. Since our pipeline only looks at frequencies greater than $5~d^{-1}$, the peak which is picked as $\nu_0$ is about $\nu=9~d^{-1}$. Based on PLR in the bottom-right panel, this object would be considered a contaminant. However, if we were to place this star at the higher frequency peaks, the object would be classified as a $\delta$ Scuti star. Therefore, this object is classified as a false-negative and a hybrid pulsator. This makes sense when considering that the star is along the red edge of the instability strip in the bottom-left panel.

Having analyzed 200 objects, we find 84 TPs, 101 TNs, 12 FNs, and 3 FPs. We additionally find that hybrids are numerous in this sample. This highlights the value of a $5~d^{-1}$ lower frequency threshold adopted in this work. While $g$-mode contamination still results in 7 FNs, the bulk of $g$-modes are below our threshold, allowing for more $\delta$ Scuti stars to be identified.

\subsection{True Positives}

\begin{figure*}[ht!]
    \plotone{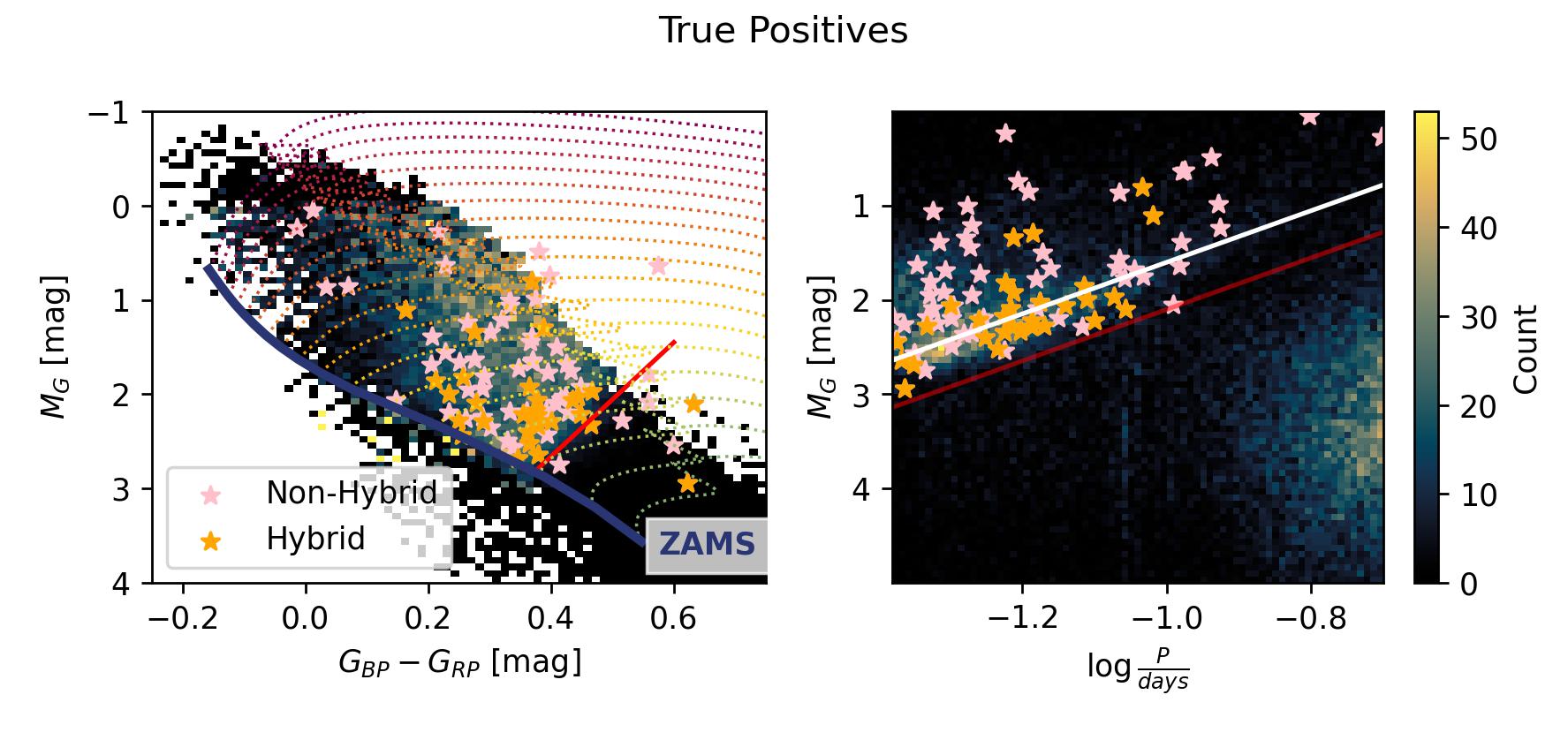}
    \caption{\textit{Left:} a reproduction of Figure \ref{fig:pulse_frac}, showing pulsator fraction as a function of color and magnitude. The edges of the instability strip are plotted in red and blue, and the location of TPs are plotted as orange stars where objects are hybrids, and pink stars otherwise. \textit{Right:} a reproduction of Figure \ref{fig:plr}, showing the density of objects as a function of log-pulsation period and absolute $G$ magnitude. Our measured PLR is plotted in white, and the line which distinguishes $\delta$ Scuti variables from contaminants is plotted in red. The location of TPs are plotted as orange stars where objects are hybrids, and pink stars otherwise.}
    \label{fig:TP}
\end{figure*}

In this classification, a TP is an object which is marked as a $\delta$ Scuti star, which pass a manual check. In the manual check we make sure that the light-curves and amplitude spectrum have no obvious irregularities, and are not eclipsing binaries. 

Since hybrid pulsators include genuine $\delta$ Scuti pulsations, hybrids are considered $\delta$ Scuti stars for the purposes of this classification. Out of the 84 TPs, 36\% are hybrids. Therefore in our sample of $\delta$ Scuti variables, a significant fraction are hybrid pulsators. These objects either have $g$-mode peaks which are between 1 and 5 $d^{-1}$, or peaks above 5 $d^{-1}$ which are smaller than the $p$-mode peaks.

The left panel of Figure \ref{fig:TP} shows that TPs are clustered within the high pulsator fraction regions of the instability strip as expected, and include many hybrid pulsators, particularly about the red edge.

\section{True Negatives}
\begin{figure*}[ht!]
    \plotone{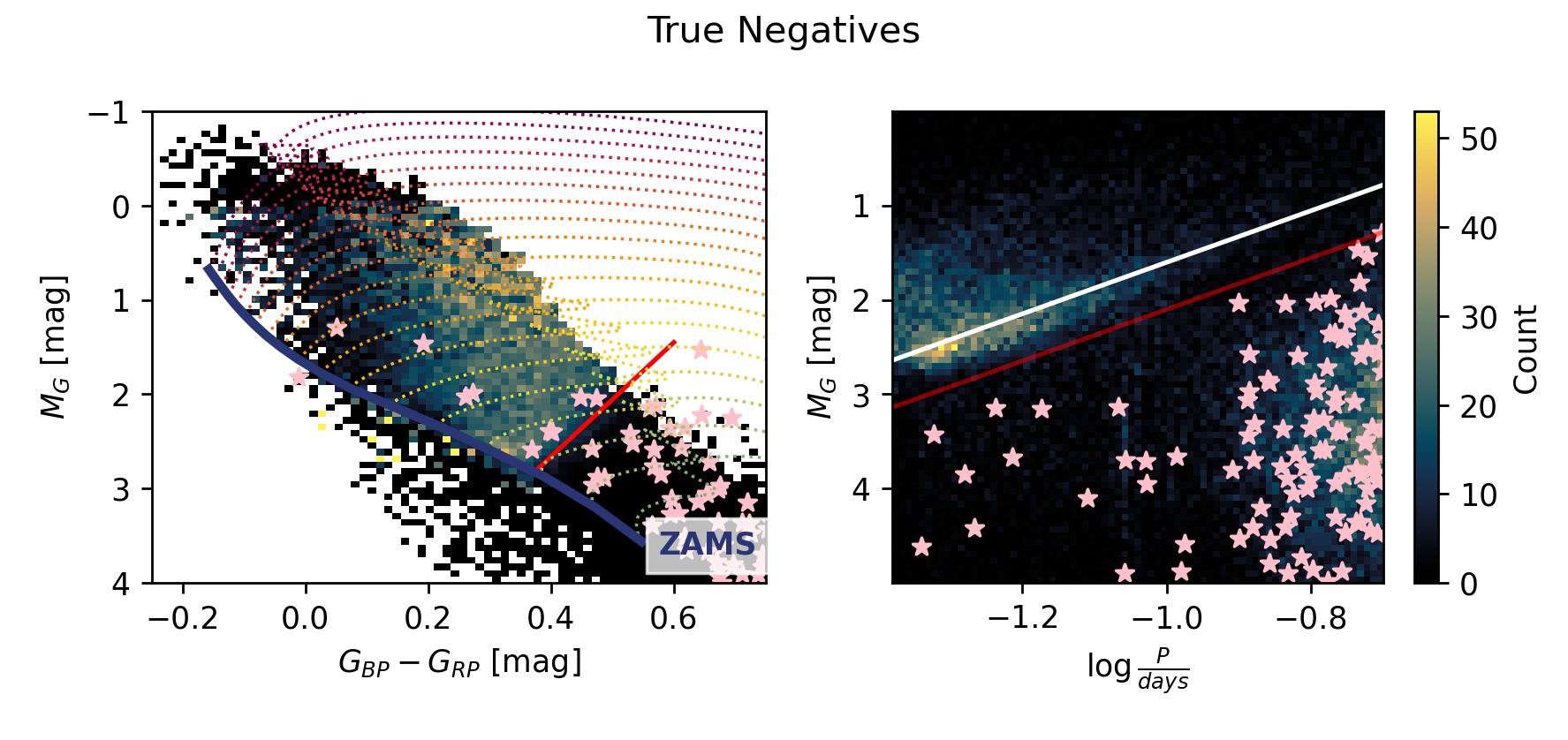}
    \caption{\textit{Left:} a reproduction of Figure \ref{fig:pulse_frac}, showing pulsator fraction as a function of color and magnitude. The edges of the instability strip are plotted in red and blue, and the location of TNs are plotted as pink stars. \textit{Right:} a reproduction of Figure \ref{fig:plr}, showing the density of objects as a function of log-pulsation period and absolute $G$ magnitude. Our measured PLR is plotted in white, and the line which distinguishes $\delta$ Scuti variables from contaminants is plotted in red. The location of TNs are plotted as pink stars.}
    \label{fig:TN}
\end{figure*}

The TNs are the most numerous of the objects we checked. To be a TN the object was marked as a variable source, but was screened out, either as an EB or via the PLR, and then passed a manual check. The manual check was based on the identified modes relation to the PLR, the objects location in the CMD in relation to the instability strip, and whether there are additional, significant $p$-mode peaks in the amplitude spectrum. Most objects in this category are clear $g$-mode pulsators, or EBs. One interesting exception was TIC 445190106, which was a RR Lyrae.

The left panel of Figure \ref{fig:TN} shows that TNs cluster to the red side of the instability strip, with some exceptions. Those hottest TNs could possibly be rapidly pulsating $\delta$ Scuti stars which have their peaks aliased into the $g$-mode island as \textbf{mentioned in the discussions of the blue side of the instability strip in \S\ref{sec:instabilitystrip} and \S\ref{sec:vbroad}}. Interestingly the upper instability strip is free of TNs. The right panel shows that TNs cluster most tightly around the $g$-mode island in the lower right of the P-L diagram, with some along the left, high frequency side.

\subsection{False Positives}
\begin{figure*}[ht!]
    \plotone{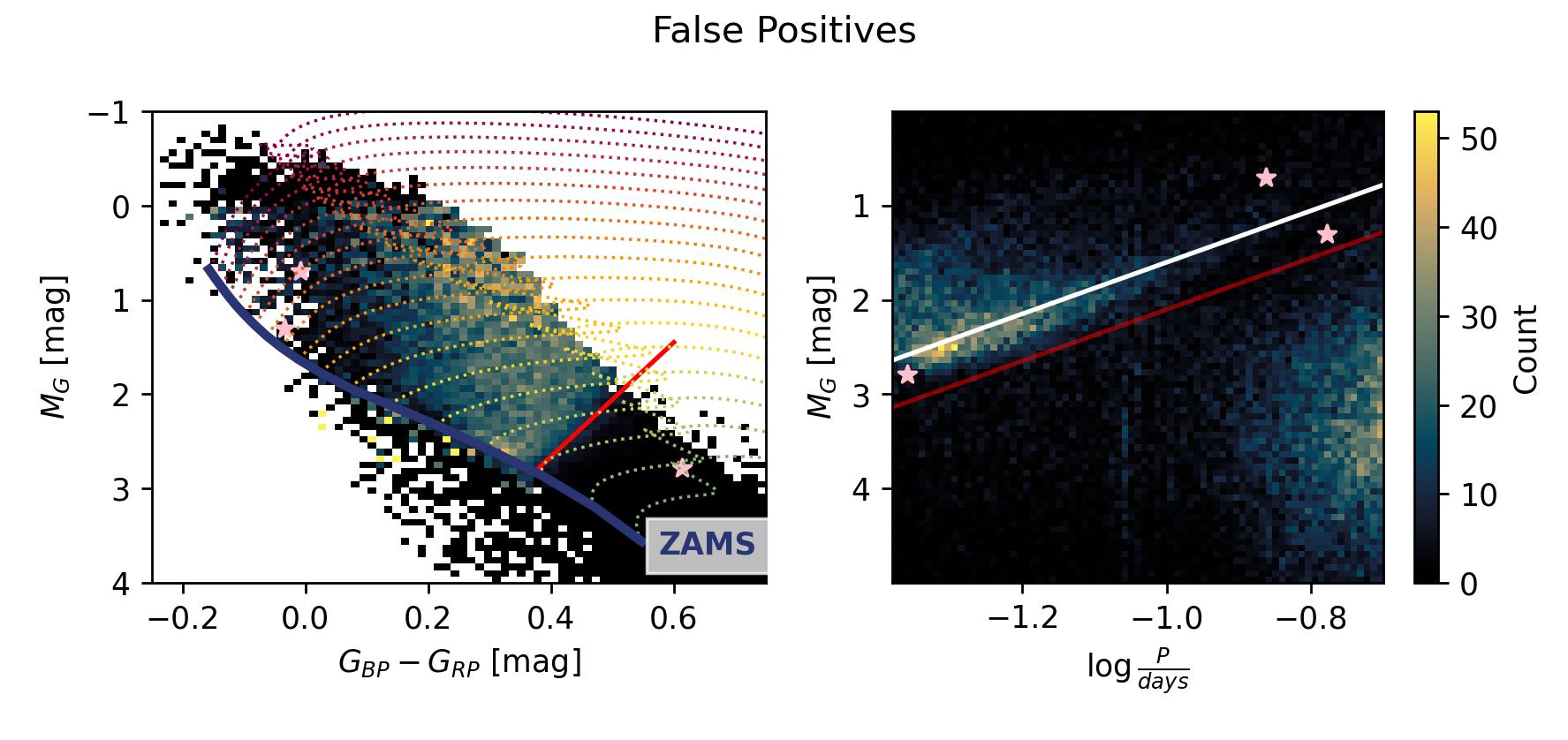}
    \caption{\textit{Left:} a reproduction of Figure \ref{fig:pulse_frac}, showing pulsator fraction as a function of color and magnitude. The edges of the instability strip are plotted in red and blue, and the location of FPs are plotted as pink stars. \textit{Right:} a reproduction of Figure \ref{fig:plr}, showing the density of objects as a function of log-pulsation period and absolute $G$ magnitude. Our measured PLR is plotted in white, and the line which distinguishes $\delta$ Scuti variables from contaminants is plotted in red. The location of FPs are plotted as pink stars.}
    \label{fig:FP}
\end{figure*}

FPs are the most concerning category of object. Much of our analysis is designed to limit the frequency of false-positives, so it is not surprising that this is the smallest group of objects. FPs were 2 misidentified EBs, and one irregular light-curve which we ascribe to poor data reduction. The EBs are both short period systems, and particularly bright, bringing them closer to the PLR.

Of the three FPs two are slowly variable, blue, and high luminosity, the other is redwards of the instability strip, low luminsoity, and rapidly variable.

\subsection{False Negatives}
\begin{figure*}[ht!]
    \plotone{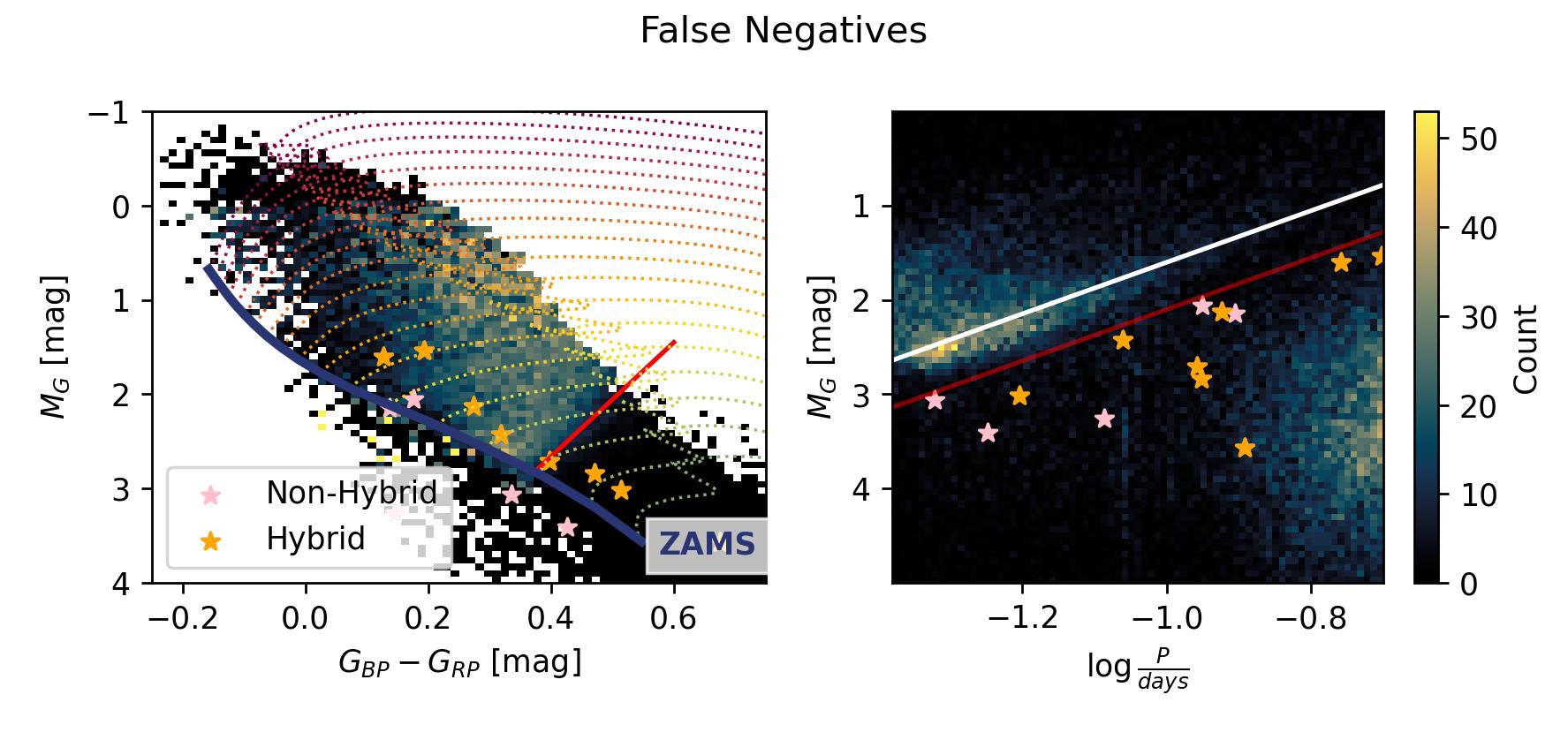}
    \caption{\textit{Left:} a reproduction of Figure \ref{fig:pulse_frac}, showing pulsator fraction as a function of color and magnitude. The edges of the instability strip are plotted in red and blue, and the location of FNs are plotted as orange stars where objects are hybrids, and pink stars otherwise. \textit{Right:} a reproduction of Figure \ref{fig:plr}, showing the density of objects as a function of log-pulsation period and absolute $G$ magnitude. Our measured PLR is plotted in white, and the line which distinguishes $\delta$ Scuti variables from contaminants is plotted in red. The location of FNs are plotted as orange stars where objects are hybrids, and pink stars otherwise.}
    \label{fig:FN}
\end{figure*}

The FNs are an interesting group of objects, and point the way for future work to improve upon these methods. The FNs were, mostly (62\%), hybrids who had their $g$-modes detected rather than their $p$-modes. In those cases, since we are only using the dominant frequency to classify objects, we are missing a significant number of stars which host $\delta$ Scuti-type pulsations. Figure \ref{fig:FN} shows that these FNs are spread along the main-sequence, and are mostly closer to our measured PLR than to the $g$-mode island in the P-L diagram.

We note that since we only use 30-minute cadence light-curves, this analysis is still susceptible to Nyquist aliasing. For this reason it is difficult to tell the difference between a true and a false negative. Since super-Nyquist signals are attenuated, we are most likely to detect quickly pulsating $\delta$ Scuti stars near to the high frequency limit, further away from the $g$-mode island. From this one could assume that all of the TNs outside of the $g$-mode island are simply super-Nyquist$\delta$ Scuti pulsators. In this analysis, aside from when an object is obviously a hybrid, we classify a negative as a TN when the star is well outside of the instability strip, and a FN when the star is inside of the instability strip. For this reason, all of the TNs with $\log \frac{P}{days} < -1$ are clustered at the far red side of the color magnitude diagram. 

\startlongtable
\begin{center}
\begin{deluxetable*}{l|cccccccc}  
    \tablecolumns{9}
    \setlength{\tabcolsep}{2.9pt}
    \tablecaption{Manually Classified Stars}
    \tablehead{
        \colhead{\emph{TESS} ID} & 
        \colhead{Classification} & 
        \colhead{Automated EB} &
        \colhead{Manual EB} &
        \colhead{$\delta$ Scuti} &
        \colhead{Hybrid} &
        \colhead{$\nu_0$} &
        \colhead{$A_0$} &
        \colhead{$N_{harmonics}$} 
    }
    \startdata
        162694156 & tp & 0 & 0 & 1 & 0 & 19.11 & 505.1 & ---\\
        172309348 & fn & 0 & 0 & 0 & 1 & 9.069 & 1207 & 0\\
        406118886 & tn & 0 & 1 & 0 & 0 & 7.547 & 7356 & ---\\
        459769845 & tp & 0 & 0 & 1 & 0 & 17.93 & 1518 & 0\\
        343723235 & tp & 0 & 0 & 1 & 1 & 13.81 & 698.2 & 0\\
        396142911 & tp & 0 & 0 & 1 & 0 & 18.87 & 4053 & 0\\
        122258063 & tp & 0 & 0 & 1 & 1 & 14.76 & 259.5 & 0\\
        281741839 & tp & 0 & 0 & 1 & 0 & 21.28 & 206.6 & 0\\
        28690048 & tp & 0 & 0 & 1 & 1 & 14.98 & 1414 & ---\\
        219158142 & tp & 0 & 0 & 1 & 1 & 21.49 & 840.0 & 0\\
        352418977 & tn & 0 & 0 & 0 & 0 & 5.384 & 387.1 & ---\\
        440869145 & tp & 0 & 0 & 1 & 0 & 14.47 & 456.5 & 0\\
        121204172 & tp & 0 & 0 & 1 & 0 & 18.14 & 2752 & 0\\
        7418085 & tn & 0 & 0 & 0 & 0 & 5.645 & 130.7 & ---\\
        155942413 & tn & 0 & 0 & 0 & 0 & 7.409 & 134.4 & ---\\
        48334365 & tp & 0 & 0 & 1 & 1 & 15.35 & 1104 & 1\\
        469028941 & tp & 0 & 0 & --- & 1 & 15.39 & 603.5 & 0\\
        12630848 & tp & 0 & 0 & 1 & 1 & 10.44 & 369.9 & 0\\
        177825912 & tn & 0 & 1 & 0 & 0 & 5.583 & 1089.8 & ---\\
        447536953 & tn & 0 & 0 & --- & 0 & 5.835 & 367.0 & ---\\
        457755081 & tp & 0 & 0 & 1 & 0 & 15.13 & 643.4 & 0\\
        229401458 & tn & 0 & 0 & 0 & 0 & 7.726 & 89.42 & ---\\
        136303530 & tn & 0 & 0 & 0 & 0 & 5.465 & 570.0 & ---\\
        349902873 & fn & 0 & 0 & 0 & 1 & 5.041 & 418.2 & 0\\
        436408784 & fn & 0 & 0 & 0 & 1 & 8.935 & 3442 & 0\\
        297963957 & tp & 0 & 0 & 1 & 0 & 8.473 & 6108 & 0\\
        236455973 & tp & 0 & 0 & 1 & 0 & 9.557 & 1710.0 & 0\\
        259810544 & tp & 0 & 0 & 1 & 1 & 15.94 & 526.9 & 0\\
        51514060 & tp & 0 & 0 & 1 & 0 & 14.85 & 80.18 & 0\\
        97486662 & tn & 0 & 0 & 0 & 0 & 6.383 & 273.3 & ---\\
        460604103 & tp & 0 & 0 & 1 & 0 & 21.62 & 798.9 & 0\\
        79379680 & tn & 0 & 0 & 0 & 0 & 17.24 & 139.5 & ---\\
        164530781 & tp & 0 & 0 & 1 & 0 & 8.422 & 1552 & 0\\
        334403759 & tn & 0 & 0 & 0 & 0 & 12.87 & 401.4 & ---\\
        322645853 & fn & 0 & 0 & 0 & 0 & 8.043 & 320.6 & 0\\
        34470689 & tn & 0 & 0 & --- & 0 & 5.659 & 120.4 & ---\\
        376950829 & tn & 0 & 0 & 0 & 0 & 6.056 & 407.2 & ---\\
        282347382 & tp & 0 & 0 & 1 & 0 & 6.347 & 17.00 & 0\\
        266688542 & tn & 0 & 0 & 0 & 0 & 5.033 & 364.9 & ---\\
        173139598 & tp & 0 & 0 & 1 & 0 & 8.667 & 1223.7 & 0\\
        267832977 & tp & 0 & 0 & 1 & 1 & 16.37 & 2325 & ---\\
        152308886 & tp & 0 & 0 & 1 & 1 & 10.80 & 1013.1 & 0\\
        182887082 & tn & 0 & 0 & 0 & 0 & 6.783 & 440.2 & ---\\
        157583659 & tp & 0 & 0 & 1 & 0 & 15.56 & 103.3 & ---\\
        346465288 & tn & 0 & 0 & 0 & 0 & 5.435 & 465.4 & ---\\
        400674143 & tp & 0 & 0 & 1 & 0 & 20.36 & 175.5 & 0\\
        368717500 & tp & 0 & 0 & 1 & 1 & 16.46 & 1630 & 0\\
        409646389 & fn & 0 & 0 & 0 & 0 & 17.72 & 1111 & 0\\
        42714007 & tp & 0 & 0 & 1 & 0 & 20.26 & 892.3 & 0\\
        394731585 & tn & 0 & 1 & 0 & 0 & 5.393 & 146.6 & ---\\
        316471937 & fn & 0 & 0 & 0 & 1 & 11.49 & 95.38 & ---\\
        275502109 & tn & 0 & 0 & --- & 0 & 11.42 & 225.1 & ---\\
        342673240 & tp & 0 & 0 & 1 & 0 & 21.47 & 898.6 & 0\\
        159347992 & tn & 0 & 0 & 0 & 0 & 7.681 & 197.4 & ---\\
        165417347 & tp & 0 & 0 & 1 & 0 & 11.45 & 1829 & 0\\
        429411041 & tn & 0 & 0 & 0 & 0 & 6.204 & 319.8 & ---\\
        40593805 & tn & 0 & 0 & --- & 0 & 7.684 & 3131 & ---\\
        335994330 & tn & 0 & 0 & 0 & 0 & 9.679 & 151.6 & ---\\
        376281533 & tp & 0 & 0 & 1 & 1 & 22.40 & 376.1 & 0\\
        177408648 & fn & 0 & 0 & 0 & 1 & 8.396 & 446.3 & 0\\
        110086948 & tn & 0 & 0 & 0 & 0 & 6.271 & 678.1 & ---\\
        396531885 & fp & 0 & 0 & 1 & 0 & 22.89 & 552.5 & 0\\
        267897033 & tn & 0 & 1 & 0 & 0 & 6.215 & 2832.0 & ---\\
        326787473 & tn & 0 & 0 & 0 & 0 & 11.41 & 100.5 & ---\\
        311178370 & tp & 0 & 0 & 1 & 1 & 12.92 & 1505 & 0\\
        119580776 & tn & 0 & 0 & 0 & 0 & 5.835 & 436.5 & ---\\
        396903512 & fp & 0 & 1 & --- & 0 & 5.993 & 4065.1 & 0\\
        201114459 & tn & 0 & 0 & 0 & 0 & 7.192 & 623.3 & ---\\
        351548657 & tn & 0 & 0 & 0 & 0 & 6.620 & 129.6 & ---\\
        369027402 & tn & 0 & 0 & 0 & 0 & 5.827 & 223.2 & ---\\
        366973356 & tn & 0 & 0 & 0 & 0 & 6.904 & 401.0 & ---\\
        7597696 & tn & 0 & 0 & 0 & 0 & 7.955 & 801.3 & 0\\
        415616799 & tn & 0 & 0 & 0 & 0 & 5.979 & 326.7 & ---\\
        124510028 & tn & 0 & 1 & 0 & 0 & 5.275 & 909.2 & ---\\
        327395181 & tp & 0 & 0 & 1 & 0 & 23.92 & 485.2 & 0\\
        93853127 & tn & 0 & 0 & 0 & 0 & 16.35 & 209.8 & ---\\
        468979699 & tp & 0 & 0 & 1 & 1 & 11.84 & 1009 & 1\\
        365601028 & tp & 0 & 0 & 1 & 0 & 11.09 & 1911 & 0\\
        118408851 & fn & 0 & 0 & 0 & 0 & 8.925 & 256.6 & 0\\
        95514282 & tn & 0 & 0 & 0 & 0 & 5.094 & 154.1 & ---\\
        83059647 & fn & 0 & 0 & 0 & 1 & 15.99 & 1024 & 0\\
        137084042 & tp & 0 & 0 & 1 & 0 & 9.516 & 4911 & 0\\
        331818226 & fn & 0 & 0 & 0 & 1 & 5.734 & 134.6 & ---\\
        423391498 & tp & 0 & 0 & 1 & 0 & 11.73 & 3000 & 0\\
        436660518 & tn & 0 & 1 & 0 & 0 & 5.849 & 2292 & ---\\
        48188920 & tp & 0 & 0 & 1 & 0 & 16.79 & 352.6 & 0\\
        441801911 & tp & 0 & 0 & 1 & 0 & 22.19 & 868.6 & ---\\
        307035635 & tp & 0 & 0 & 1 & 1 & 23.63 & 712.3 & 0\\
        162090465 & tn & 0 & 0 & 0 & 0 & 6.125 & 448.9 & ---\\
        130416157 & fp & 0 & 0 & 1 & 0 & 7.286 & 26.15 & ---\\
        260935220 & tn & 0 & 0 & --- & 0 & 5.925 & 1621 & ---\\
        90033607 & tn & 0 & 0 & 0 & 0 & 5.129 & 235.2 & 0\\
        143614026 & tn & 0 & 0 & --- & 0 & 14.88 & 823.1 & ---\\
        80309353 & tp & 0 & 0 & 1 & 0 & 5.029 & 3362 & 0\\
        411468522 & tn & 0 & 0 & 0 & 0 & 5.317 & 407.9 & ---\\
        165060131 & tn & 0 & 1 & 0 & 0 & 7.578 & 947.0 & ---\\
        20095466 & tn & 0 & 1 & 0 & 0 & 9.439 & 301.6 & 1\\
        2003348372 & tn & 0 & 0 & 0 & 0 & 6.839 & 333.7 & ---\\
        406637709 & tn & 0 & 0 & 0 & 0 & 7.704 & 858.6 & ---\\
        233574062 & tn & 0 & 0 & 0 & 0 & 6.678 & 5545.8 & ---\\
        85000324 & tn & 0 & 0 & 0 & 0 & 5.352 & 406.7 & ---\\
        443212084 & tp & 0 & 0 & 1 & 1 & 16.74 & 1564 & 0\\
        53558656 & tp & 0 & 0 & 1 & 1 & 16.32 & 798.0 & 0\\
        155167702 & tp & 0 & 0 & 1 & 0 & 14.99 & 974.4 & 0\\
        219923095 & tp & 0 & 0 & 1 & 1 & 17.85 & 913.7 & 1\\
        272561697 & tp & 0 & 0 & 1 & 0 & 21.09 & 283.0 & 0\\
        85901731 & tp & 0 & 0 & 1 & 0 & 15.17 & 105.9 & ---\\
        295380484 & fn & 0 & 0 & 0 & 0 & 12.18 & 557.6 & ---\\
        144177328 & tn & 0 & 0 & 0 & 0 & 7.918 & 627.9 & ---\\
        456311127 & fn & 0 & 0 & 0 & 0 & 20.97 & 3018 & 0\\
        240982077 & tn & 0 & 1 & 0 & 0 & 6.574 & 2853 & ---\\
        394579246 & tn & 0 & 0 & 0 & 0 & 6.245 & 258.2 & ---\\
        188572955 & tn & 0 & 0 & 0 & 0 & 5.158 & 157.9 & ---\\
        154847435 & tn & 0 & 0 & 0 & 0 & 6.033 & 2997 & ---\\
        151704157 & tp & 0 & 0 & 1 & 0 & 19.78 & 335.7 & 0\\
        316489984 & tn & 0 & 0 & 0 & 0 & 5.942 & 102.6 & ---\\
        178666862 & tp & 0 & 0 & 1 & 0 & 9.779 & 1745 & 0\\
        273000951 & tp & 0 & 0 & 1 & 0 & 21.10 & 276.6 & 0\\
        129011402 & tp & 0 & 0 & 1 & 1 & 16.78 & 1227 & 0\\
        32530825 & tp & 0 & 0 & 1 & 1 & 16.29 & 1157 & 0\\
        190180185 & tn & 0 & 0 & 0 & 0 & 5.199 & 850.4 & ---\\
        305967690 & tn & 0 & 0 & 0 & 0 & 5.990 & 78.33 & ---\\
        445190106 & tn & 0 & 1 & 0 & 0 & 6.327 & 1399.4 & 1\\
        286523081 & tp & 0 & 0 & 1 & 0 & 18.65 & 846.0 & 0\\
        233417705 & tn & 0 & 1 & 0 & 0 & 5.840 & 101.4 & 0\\
        110993631 & tn & 0 & 0 & 0 & 0 & 5.106 & 759.9 & ---\\
        290907664 & tn & 0 & 0 & 0 & 0 & 5.219 & 298.1 & ---\\
        250308237 & tp & 0 & 0 & 1 & 0 & 18.72 & 1328 & 0\\
        279431011 & tn & 0 & 0 & 0 & 0 & 5.028 & 169.4 & ---\\
        42387893 & tn & 0 & 0 & 0 & 0 & 5.715 & 403.8 & 0\\
        341034233 & tp & 0 & 0 & 0 & 0 & 11.63 & 61.86 & ---\\
        354948311 & tn & 0 & 0 & 0 & 0 & 9.546 & 1251 & ---\\
        364354624 & tn & 0 & 1 & 0 & 0 & 5.292 & 234.2 & ---\\
        30727674 & tn & 0 & 0 & 0 & 0 & 5.088 & 792.5 & ---\\
        126068461 & tp & 0 & 0 & 1 & 0 & 10.77 & 2029.0 & 1\\
        206539678 & tp & 0 & 0 & 1 & 1 & 12.58 & 1832 & 0\\
        2846517 & tn & 0 & 0 & 0 & 0 & 6.801 & 468.5 & ---\\
        142327859 & tn & 0 & 1 & 0 & 0 & 7.178 & 1546 & ---\\
        137835456 & tn & 0 & 0 & 0 & 0 & 5.262 & 37.94 & ---\\
        311121344 & tn & 0 & 0 & 0 & 0 & 7.601 & 282.7 & ---\\
        282088971 & tn & 0 & 0 & 0 & 0 & 6.837 & 1183 & 0\\
        292468913 & tp & 0 & 0 & 1 & 0 & 9.601 & 5258 & 0\\
        429306233 & tp & 0 & 0 & 1 & 0 & 16.73 & 111.3 & 0\\
        454506804 & tp & 0 & 0 & 1 & 1 & 23.08 & 903.2 & 0\\
        271779389 & tp & 0 & 0 & 1 & 0 & 18.61 & 1126 & 0\\
        47513732 & tn & 0 & 0 & 0 & 0 & 5.734 & 98.42 & ---\\
        256750557 & tn & 0 & 0 & 0 & 0 & 5.108 & 2003 & ---\\
        33599701 & tn & 0 & 0 & 0 & 0 & 7.206 & 223.4 & ---\\
        56138466 & tp & 0 & 0 & 1 & 1 & 13.02 & 123.6 & 0\\
        449733031 & tn & 0 & 0 & 0 & 0 & 19.04 & 78.78 & ---\\
        407166550 & tp & 0 & 0 & 0 & 0 & 19.85 & 415.6 & 0\\
        314214193 & tn & 0 & 0 & 0 & 0 & 18.44 & 213.1 & ---\\
        132775655 & tp & 0 & 0 & 1 & 1 & 15.36 & 1386 & 0\\
        434456524 & tp & 0 & 0 & 1 & 1 & 17.18 & 2375 & 0\\
        303994777 & tp & 0 & 0 & 1 & 0 & 23.14 & 228.6 & ---\\
        178995864 & tn & 0 & 1 & 0 & 0 & 5.805 & 1446.7 & ---\\
        173439709 & fn & 0 & 0 & 0 & 1 & 7.799 & 2388 & 0\\
        305777433 & tp & 0 & 0 & 1 & 0 & 11.62 & 472.7 & 0\\
        299309578 & tn & 0 & 0 & 0 & 0 & 7.246 & 405.4 & ---\\
        280020045 & tp & 0 & 0 & 1 & 0 & 18.62 & 2050 & 0\\
        229798416 & tp & 0 & 0 & 1 & 0 & 9.446 & 70.66 & ---\\
        252147032 & tp & 0 & 0 & 1 & 1 & 11.41 & 2318 & 0\\
        105707877 & tn & 0 & 0 & 0 & 0 & 8.098 & 202.9 & 0\\
        282209645 & tp & 0 & 0 & 1 & 0 & 22.37 & 920.7 & 0\\
        361566051 & tn & 0 & 1 & 0 & 0 & 5.004 & 1059 & 0\\
        318966934 & tp & 0 & 0 & 1 & 0 & 14.11 & 475.4 & 0\\
        468987768 & tp & 0 & 0 & 1 & 0 & 23.75 & 461.5 & 0\\
        220183458 & tn & 0 & 0 & 0 & 0 & 10.70 & 2722 & 0\\
        377197082 & tn & 0 & 0 & 0 & 0 & 11.65 & 498.5 & ---\\
        26992387 & tn & 0 & 0 & 0 & 0 & 21.89 & 622.3 & ---\\
        362988825 & tp & 0 & 0 & 0 & 0 & 20.65 & 746.4 & 0\\
        201692249 & tn & 0 & 0 & 0 & 0 & 5.717 & 181.1 & ---\\
        79789474 & tn & 0 & 0 & 0 & 0 & 20.99 & 222.0 & ---\\
        307244763 & tn & 0 & 1 & 0 & 0 & 5.385 & 6147 & 0\\
        236965396 & tn & 0 & 0 & 0 & 0 & 6.327 & 141.1 & ---\\
        115551110 & tp & 0 & 0 & 1 & 1 & 18.21 & 1398 & 0\\
        268041302 & tn & 0 & 1 & 0 & 0 & 5.496 & 738.4 & ---\\
        331004760 & tn & 0 & 0 & 0 & 0 & 5.215 & 423.6 & ---\\
        187781811 & tn & 0 & 0 & 0 & 0 & 6.448 & 345.3 & ---\\
        373030663 & tn & 0 & 1 & 0 & 0 & 6.938 & 1290 & ---\\
        77738519 & tn & 0 & 0 & 0 & 0 & 5.658 & 2039 & ---\\
        292122717 & tn & 0 & 1 & 0 & 0 & 6.246 & 1813.6 & 1\\
        262832737 & tn & 0 & 0 & 0 & 0 & 5.011 & 661.0 & ---\\
        179295308 & tp & 0 & 0 & 1 & 0 & 16.10 & 1990 & 0\\
        450096611 & tn & 0 & 0 & 0 & 0 & 5.303 & 7344.6 & ---\\
        105534100 & tn & 0 & 1 & 0 & 0 & 5.265 & 2853 & ---\\
        423395903 & tn & 0 & 0 & 0 & 0 & 6.728 & 5578.2 & 1\\
        248997990 & tp & 0 & 0 & 1 & 0 & 19.86 & 478.1 & 0\\
        198181552 & tn & 0 & 0 & 0 & 0 & 5.669 & 59.15 & ---\\
        332915772 & tp & 0 & 0 & 1 & 1 & 16.80 & 258.0 & ---\\
        194147049 & tn & 0 & 0 & 0 & 0 & 10.64 & 541.2 & 0\\
        451222707 & tp & 0 & 0 & 1 & 1 & 19.95 & 361.8 & 0\\
        152461579 & tp & 0 & 0 & 1 & 0 & 19.00 & 1573 & 0\\
        150268967 & tp & 0 & 0 & 1 & 0 & 20.77 & 259.8 & 0\\
        392482957 & tp & 0 & 0 & 1 & 1 & 23.30 & 1149 & ---\\
        163147809 & tp & 0 & 0 & 1 & 0 & 13.11 & 4017 & 0\\
        82575550 & tn & 0 & 0 & 0 & 0 & 5.408 & 508.8 & ---\\
        13068361 & tn & 0 & 1 & 0 & 0 & 6.508 & 1085.93 & ---\\
        37657836 & tp & 0 & 0 & 1 & 0 & 21.31 & 130.0 & 0\\
        267751030 & tn & 0 & 1 & 0 & 0 & 5.455 & 9793 & ---
    \enddata
\end{deluxetable*}
\end{center}

\end{document}